 \documentclass[preprint,12pt]{elsarticle}

\usepackage{amssymb}
\usepackage{amsmath}
\usepackage{bm}
\usepackage{physics}
\usepackage{tikz}

\newcommand{\N}{\mathcal{N}_{\alpha}}
\newcommand{\nN}{\mathcal{\hat{N}}_\alpha}
\renewcommand{\a}{\alpha}

\newcommand{\csref}{c_{\mathrm{ref}}}
\newcommand{\nref}{n_{\mathrm{ref}}}
\newcommand{\Lref}{L_{\mathrm{ref}}}
\newcommand{\rhoref}{\rho_{\mathrm{ref}}}
\newcommand{\Bref}{B_{\mathrm{ref}}}
\newcommand{\mref}{m_{\mathrm{ref}}}

\newcommand{\Tref}{T_{\mathrm{ref}}}
\newcommand{\Omegaref}{\Omega_{\mathrm{ref}}}
\newcommand{\betaref}{\beta_{\mathrm{ref}}}
\renewcommand{\pb}[2]{\left\{ #1, #2\right\}}
\newcommand{\curv}[1]{C\left( #1\right)}

\journal{Journal of Computational Physics}

\begin{document}

\begin{frontmatter}

%% Title, authors and addresses

%% use the tnoteref command within \title for footnotes;
%% use the tnotetext command for theassociated footnote;
%% use the fnref command within \author or \affiliation for footnotes;
%% use the fntext command for theassociated footnote;
%% use the corref command within \author for corresponding author footnotes;
%% use the cortext command for theassociated footnote;
%% use the ead command for the email address,
%% and the form \ead[url] for the home page:
%% \title{Title\tnoteref{label1}}
%% \tnotetext[label1]{}
%% \author{Name\corref{cor1}\fnref{label2}}
%% \ead{email address}
%% \ead[url]{home page}
%% \fntext[label2]{}
%% \cortext[cor1]{}
%% \affiliation{organization={},
%%             addressline={},
%%             city={},
%%             postcode={},
%%             state={},
%%             country={}}
%% \fntext[label3]{}

\title{Spectrally Accelerated Edge and Scrape-Off Layer Gyrokinetic Turbulence Simulations}

%% use optional labels to link authors explicitly to addresses:
%% \author[label1,label2]{}
%% \affiliation[label1]{organization={},
%%             addressline={},
%%             city={},
%%             postcode={},
%%             state={},
%%             country={}}
%%
%% \affiliation[label2]{organization={},
%%             addressline={},
%%             city={},
%%             postcode={},
%%             state={},
%%             country={}}

\author[ipp]{B. J. Frei} %% Author name
\author[ipp]{P. Ulbl} 
\author[ipp]{J. Trilaksono}
\author[ipp,austin]{F. Jenko}

%% Author affiliation
\affiliation[ipp]{organization={Max-Planck Institute for Plasma Physics},
            addressline={Boltzmannstr. 2}, 
            city={Garching},
            postcode={D-85748}, 
            country={Germany}}
            
\affiliation[austin]{organization={Institute for Fusion Studies, The University of Texas at Austin}, 
            city={Austin},
            postcode={TX 78712}, 
            country={USA}}

%% Abstract
\begin{abstract}
This paper presents the first gyrokinetic (GK) simulations of edge and scrape-off layer (SOL) turbulence accelerated by a velocity-space spectral approach in the full-$f$ GK code \verb|GENE-X|.
 Building upon the original grid velocity-space discretization, we derive and implement a new spectral formulation and verify the numerical implementation using the method of manufactured solution. We conduct a series of spectral turbulence simulations focusing on the TCV-X21 reference case [Oliveira D. S. \textit{et al.}, Nucl. Fusion \textbf{62}, 096001 (2022)] and compare these results with previously validated grid simulations [Ulbl P. \textit{et al.}, Phys. Plasmas \textbf{30}, 107986 (2023)]. This shows that the spectral approach reproduces the outboard midplane (OMP) profiles (density, temperature, and radial electric field), dominated by trapped electron mode (TEM) turbulence, with excellent agreement and significantly lower velocity-space resolution. Thus, the spectral approach reduces the computational cost by at least an order of magnitude, achieving a speed-up of approximately $50$ for the TCV-X21 case. This enables high-fidelity GK simulations to be performed within a few days on modern CPU-based supercomputers for medium-sized devices and establishes \verb|GENE-X| as a powerful tool for studying edge and SOL turbulence, moving towards reactor-relevant devices like ITER.
\end{abstract}

%%Graphical abstract
%\begin{graphicalabstract}
%\includegraphics{grabs}
%\end{graphicalabstract}

%%Research highlights
%\begin{highlights}
%\item Highlight 1
%\end{highlights}

%% Keywords
%\begin{keyword}
%Plasma, Gyrokinetic, Turbulence
%% keywords here, in the form: keyword \sep keyword

%% PACS codes here, in the form: \PACS code \sep code

%% MSC codes here, in the form: \MSC code \sep code
%% or \MSC[2008] code \sep code (2000 is the default)

%\end{keyword}

\end{frontmatter}

%% Add \usepackage{lineno} before \begin{document} and uncomment 
%% following line to enable line numbers
%% \linenumbers

%% main text
%%

%% Use \section commands to start a section
\section{Introduction}

Predicting turbulent transport in the edge and SOL regions is crucial to optimize fusion reactor performance, predict the divertor heat load, understand the L- to H-mode confinement transition, and design future magnetic confinement fusion devices, such as ITER \cite{ikeda2007} and DEMO \cite{zohm2013}. Despite significant recent advancements in turbulent transport modeling \cite{litaudon2022}, widely-used reduced turbulent transport (e.g., quasilinear) models and Braginskii-fluid simulations often fail to accurately capture edge and SOL turbulent transport due to, for instance, the importance of non-local effects and the fact that the edge is only marginally collisional. Therefore, because of the peculiar properties of the edge and SOL region, high-fidelity gyrokinetic (GK) turbulence simulations are necessary to overcome these difficulties and describe turbulent transport accurately. 

While global \cite{goerler2011,grandgirard2016,lanti2020} and local \cite{jenko2000,candy2016,peeters2009} GK codes for core turbulence are well-established, GK turbulence codes for the edge and SOL region remain less mature. One of the main reasons for this is (i) the lack of a clear separation between fluctuations and equilibrium quantities requiring a full-$f$ GK formalism and (ii) the complex magnetic geometry featuring open and closed field lines and X-points, which poses significant numerical challenges. In addressing this latter complexity, Braginskii-fluid turbulence codes such as \verb|GBS| \cite{giacomin2021}, \verb|TOKAM3X| \cite{tamain2016},  and \verb|GRILLIX| \cite{stegmeir2019} have pioneered SOL turbulence, where the high-collisional assumption might be justified, in arbitrary magnetic configurations. Notably, the flux-coordinate independent (FCI) approach \cite{hariri2013,stegmeir2016}, implemented in the \verb|GRILLIX| fluid code \cite{stegmeir2019, zholobenko2021}, has demonstrated promising performance in simulating turbulence in large devices \cite{stegmeir2023}, with flexible magnetic geometries \cite{body2020}.

Building on the experience gained from the \verb|GENE| code \cite{jenko2000,goerler2011} and the flexibility of the FCI method, the \verb|GENE-X| code \cite{michels2021} has been specifically designed to perform high-fidelity and high-performance GK turbulence simulations of the edge and SOL region with X-points. \verb|GENE-X| is a full-$f$ GK code, i.e., it does not split the distribution function between an equilibrium and fluctuating parts. More precisely, \verb|GENE-X| solves the full-$f$ electromagnetic and collisional GK Vlasov-Maxwell system and belongs to the continuum category of GK codes, where an Eulerian grid approach is utilized to discretize the velocity-space. Currently, \verb|GENE-X| is one of the few GK codes able to perform edge and SOL turbulence with magnetic X-points. Among the other existing full-$f$ GK codes designed for edge and SOL applications, we can cite \verb|GKEYLL| \cite{shi2017} and \verb|PICLS| \cite{boesl2019}, which focus on either open and closed the open-field line region, and \verb|XGC| \cite{ku2009, hager2022} and \verb|COGENT| \cite{dorf2021}, which can also include magnetic X-points. 

The \verb|GENE-X| code has been validated against attached L-mode experiments in medium-sized devices such as in the ASDEX Upgrade \cite{michels2022} and TCV tokamak \cite{ulbl2023}. These validations demonstrate that \verb|GENE-X| can provide predictions close to experimental measurements (e.g., OMP profiles, power balance, and divertor fall of length) in L-mode conditions. For instance, in the TCV-X21 reference case (L-mode discharge designed for code validation \cite{oliveira2022}), the \verb|GENE-X| simulations have revealed the importance of the collisional cooling of trapped electrons in the edge to recover the correct electron temperature OMP profile within the experimental uncertainty, a kinetic mechanism that Braginskii-like fluid codes fail to capture due to the absence of trapped particles in these models. Despite these promising and encouraging results, the significant computational requirements of these first-principles grid GK simulations hinders the ability of \verb|GENE-X| to simulate edge and SOL turbulence in reactor-relevant devices (such as ITER) and to explore high-performance and advanced experimental scenarios. Even for medium-sized devices, simulations frequently require several million CPU hours and span over several weeks to complete \cite{michels2022, ulbl2023}. Although GPUs and exascale HPC architectures offer potential, new numerical algorithms are needed to accelerate high-fidelity GK simulations. 

This paper presents the first implementation of a spectral approach in velocity-space in a full-$f$ GK turbulence code such as \verb|GENE-X|. The use of a spectral method is motivated by the fact that it can be particularly advantageous at high collisionality (e.g., in the SOL). The spectral method used in this work is based on a spectral expansion of the full-$f$ distribution function onto a Hermite and Laguerre polynomial basis in velocity-space. Using this basis, we derive and numerically implement the spectral formulation of the edge and SOL GK turbulence model solved in \verb|GENE-X|. The numerical implementation is verified using the method of manufactured solution (MMS) \cite{roache2002}. 
We present the first spectrally accelerated GK edge and SOL turbulence simulations of the TCV-X21 reference case using \verb|GENE-X|. It is noteworthy that the TCV-X21 scenario represents an ideal reference case for assessing the performance of the spectral approach, given that turbulence is dominated by TEMs \cite{ulbl2023}, which can present a challenge for a global velocity-space spectral approach \cite{frei2023}. We compare our spectral results with the ones of Ref. \cite{ulbl2023} obtained from grid simulations with \verb|GENE-X|. We find that the spectral approach can reproduce the OMP profiles in both the collisional and collisionless cases, dominated by TEMs, with excellent agreement and with a small number of spectral coefficients. In addition, by further increasing the spectral resolution in our simulations, we demonstrate that the agreement between the grid and spectral results is improved. Finally, the computational cost of the spectral simulations is assessed. We demonstrate that the spectral approach implemented in this work achieves a significant speed-up of CPU-based GK simulations with \verb|GENE-X|. For the TCV-X21 reference case, a speed-up of nearly $50$ times is achieved compared to the previous grid simulations, allowing high-fidelity GK edge and SOL turbulence simulations to be conducted within a few days on current CPU-based supercomputers. This opens up new opportunities for studying edge and SOL turbulence through high-fidelity and high-performance GK simulations, which is crucial for the success of ITER and the design of future fusion power plants.

This paper is structured as follows. First, we introduce the GK turbulence model for the edge and SOL used in \verb|GENE-X| in Section \ref{sec:genexmodel}. We then present the spectral approach considered in this work and derive the spectral formulation of the GK model in Section \ref{sec:spectralgenex}. The numerical implementation of the spectral approach in \verb|GENE-X| is detailed in Section \ref{sec:numericalimplementation}, while its verification is carried out in Section \ref{sec:mms}. The first spectrally accelerated turbulence simulations of the TCV-X21 reference case are reported in Section \ref{sec:tcvx21}, including a comparison with the grid simulations from Ref. \cite{ulbl2023}. Finally, the performance and computational cost of the spectral simulations are evaluated in Section \ref{sec:computational}. The conclusions and outlooks are presented in Section \ref{sec:conclusion}.

%% Use \subsection commands to start a subsection.

\section{The GK turbulence model for the edge and SOL}
\label{sec:genexmodel} 

\verb|GENE-X| evolves the full-$f$ gyrocenter and gyroaveraged distribution function, $ f_\alpha = f_\alpha(\bm{R}, v_\parallel, \mu,t)$ of particle species $\alpha$ in the gyrocenter phase-space described by the coordinates $(\bm{R}, v_\parallel, \mu, \theta)$. Here, $\bm{R}$ denotes the gyrocenter position related to the particle position $\bm{r}$ by the transformation $\bm{r} \simeq \bm{R} + \bm{\rho}_\alpha(\mu, \theta)$, with $\bm{\rho}_\alpha = \bm{v}_\perp(\mu, \theta)/ \Omega_\alpha$ the particle Larmor radius and $\Omega_\alpha = q_\alpha B / (c m_\alpha)$ the gyrofrequency, where $m_\alpha$ and $q_\alpha$ are species mass and charge, respectively. The velocity dependence of the gyrocenter phase-space is described by the parallel and perpendicular components of the gyrocenter velocity $\bm{v}$ relative to the equilibrium magnetic field, i.e., $v_\parallel = \bm{v} \cdot \bm{b}$ with background magnetic field unit vector $\bm{b} = \bm{B} / B$ and $ \bm v_\perp = \bm v - \bm b v_\parallel$, respectively. Finally, $\mu = m_\alpha \norm{\bm v_\perp}^2 /(2B)$ is the gyrocenter magnetic moment, and $\theta$ is the gyrocenter gyroangle, such that $\partial_\theta f_\alpha = 0$. To obtain the self-consistent evolution of $f_\alpha$, the long-wavelength electromagnetic and collisional GK Vlasov equation is used, which is given in conservative form,

\begin{align}
\label{eq:vlasov}
\frac{\partial \left(  B_{\|}^*  f_\alpha  \right)}{\partial t}  + \grad \cdot \left( B_{\|}^* \dot{\bm R} f_\alpha \right) +  \frac{\partial}{\partial v_{\|}}\left(   B_{\|}^* \dot v_\parallel f_\alpha \right) = \sum_{\beta} B_{\|}^* C_{\alpha \beta}(f_\alpha, f_\beta),
\end{align}
\\
where the conservation of the magnetic moment, $\dot \mu = 0$, is used. In Eq. (\ref{eq:vlasov}), the equations of motion are 

\begin{subequations} \label{eq:motions}
\begin{align}
\dot{\bm R} & = v_{\|}  \frac{\bm{B}^*}{B_{\|}^*}  + \frac{c}{q_\alpha B_{\|}^*} \bm{b} \times\left(\mu  \grad  B + q_\alpha  \grad  \psi_{1 \alpha}\right), \\
\dot v_\parallel & =  - \frac{\bm{B}^*}{m_\alpha B_{\|}^*} \cdot\left(\mu  \grad  B+q_\alpha \grad \psi_{1 \alpha}\right)  - \frac{q_\alpha}{m_\alpha c} \frac{\partial A_{1 \parallel} }{\partial t} ,
\end{align}
\end{subequations}
\\
with $\bm{B}^* =  \bm{B}_{T} + m_\alpha v_{\|} c \grad \times \bm{b} / q_\alpha$. Here, $\bm B_{T} = \bm B + \grad A_{1 \|} \times \bm{b}$ is the sum of the constant (in time and space) magnetic field, $\bm B$, and the small perpendicular magnetic fluctuation, $\delta \bm B_\perp \simeq \grad A_{1 \|} \times \bm{b}$, with $A_{1 \|}$ the parallel component of the perturbed magnetic vector potential. We remark that the gyrocenter Jacobian, $B_\parallel^* / m_\alpha$, is proportional to $B_{\parallel}^* = \bm b \cdot \bm B^* = B + m_\alpha v_{\|} c \bm b \cdot \grad \times \boldsymbol{b}  / q_\alpha$. The second term in $B_{\parallel}^*$ is referred to as the guiding-center correction and is proportional to the ratio of the particle Larmor radius $\rho_\alpha$ to the scale length of the equilibrium magnetic field, $L_B \sim R$ (with $R$ being the major radius of the fusion devices), i.e. it is a small correction proportional to $\rho_\alpha / L_B \ll 1$. It is worth noticing that the grid implementation of \verb|GENE-X| (see Section \ref{sec:numericalimplementation}) solves the GK Vlasov equation in the advection form rather than the conservative form given in Eq. (\ref{eq:vlasov}). However, the conservative form is more convenient to derive the spectral approach in Section \ref{sec:spectralgenex}. 

In Eq. (\ref{eq:vlasov}), we introduce the generalized potential, $\psi_{1 \alpha}$, defined by

\begin{align} \label{eq:psi}
\psi_{1 \alpha} = \phi_1 -  \frac{m_\alpha c^2}{2 q _\alpha  B^2} \left|  \grad_\perp \phi_1 \right|^2,
\end{align}
\\
which is the sum of the electrostatic potential $\phi_1$ and a second-order correction term that ensures energy consistency. Here, the perpendicular gradient is defined by $\grad_\perp = (\mathbb{I} - \bm b \bm b ) \cdot \grad$ (with $\mathbb{I}$ being the identity matrix). We remark that higher order finite Larmor radius (FLR) effects are currently neglected in the \verb|GENE-X| model.

Collisional effects are introduced through the collision operator $C_{\alpha \beta}$ on the right hand-side of Eq. (\ref{eq:vlasov}). $C_{\alpha \beta}$ is modeled by a full-$f$ and multi-species collision operator, which describes the effect of Coulomb collisions between species $\alpha$ and $\beta$. While different collision operator models are currently available in \verb|GENE-X|, we consider a long-wavelength Lernard-Bernstein/Dougherty (LBD) collision operator \cite{lenard1958,dougherty1964}. The explicit expression can be found in Ref. \cite{ulbl2022}. 

The self-consistent evolution of the electromagnetic fields, $\phi_1$ and $A_{1 \parallel}$, are obtained by the quasineutrality (QN) condition and Ampere’s law, which are 

 \begin{subequations} \label{eqs:fields}
 \begin{align}
- \grad   \cdot\left(\sum_\alpha \frac{m_\alpha c^2}{B^2} \int d W f_\alpha \grad _{\perp} \phi_1\right) & =\sum_\alpha q_\alpha \int d W f_\alpha ,  \label{eq:poisson}\\
-\Delta_{\perp} A_{1 \|}  =4 \pi \sum_\alpha \frac{q_\alpha}{c} \int d W f_\alpha v_{\|}  \label{eq:ampere},
\end{align}
\end{subequations}
\\
respectively. Here, the Laplacian operator is $\Delta_{\perp} = \grad \cdot \grad_\perp$ and the phase-space volume element is $d W = 2 \pi d \mu d v_\parallel B_\|^* / m_\alpha$. To solve for the induction part of the parallel electric field, namely $\partial_t A_{1 \parallel}$, an Ohm's law can be derived. By taking the time-derivative of the Ampere's law Eq. (\ref{eq:ampere}) and using the Vlasov equation given in Eq. (\ref{eq:vlasov}), we obtain an independent field equation for $\partial_t A_{1 \parallel}$ \cite{mandell2020},

 \begin{align}
-\left(\Delta_{\perp}+4 \pi \sum_\alpha \frac{q_\alpha^2}{m_\alpha c^2} \int d W \frac{\partial f_\alpha}{\partial v_{\|}}  \right)  \frac{\partial A_{1 \parallel} }{\partial t}  & = 4 \pi \sum_\alpha \frac{q_\alpha}{c} \int d W  \left(\frac{\partial f_\alpha}{\partial t}\right)^{\star} v_{\|} \label{eq:ohm},
\end{align}
\\
In Eq. (\ref{eq:ohm}), we introduce the notation

\begin{equation} \label{eq:staticdynamic}
\frac{\partial f_\alpha}{\partial t}=\left(\frac{\partial f_\alpha}{\partial t}\right)^{\star}+\frac{q_\alpha}{m_\alpha c} \frac{\partial f_\alpha}{\partial v_{\|}} \frac{\partial A_{1 \parallel} }{\partial t},
\end{equation}
\\
where the superscript $\star$ represents the part of the time evolution of $f_\alpha$ which is independent of $\partial_t A_{1 \parallel}$ (see Eq. (\ref{eq:vlasov})). Therefore, we refer to the first term on the right-hand side of Eq. (\ref{eq:staticdynamic}) as the static (without $\partial_t A_{1 \parallel}$) part, while the second term is referred to as the dynamic (only $\partial_t A_{1 \parallel}$) part.

Eqs. (\ref{eq:vlasov}) and (\ref{eqs:fields}), referred to as the GK Vlasov-Maxwell system, define a closed set of $5$D and time partial differential equations for the full-$f$ distribution function $f_\alpha$, the electrostatic potential $\phi_1$, the parallel component of the magnetic vector potential $A_{1 \parallel}$, and the inductive part of the parallel electric field $ \partial_t A_{1 \parallel}$. The \verb|GENE-X| code was originally designed to solve the GK Vlasov-Maxwell system using a grid approach in velocity-space \cite{michels2021}. The details of the grid formulation are discussed in Section. \ref{sec:numericalimplementation}.

Finally, the GK Vlasov-Maxwell system exactly conserves the total energy $\mathcal{E}$, which is defined by \cite{scott2021}

\begin{align} \label{eq:totalenergy}
\mathcal{E} & = \sum_\alpha \int d V \int d W \left( m_\alpha v_{\|}^2+\mu B+ q_\alpha \psi_{1 \alpha} \right) f_\alpha \nonumber \\
& + \int d V \frac{1}{8 \pi}\left|\grad_{\perp} A_{1 \|}\right|^2,
\end{align}
\\
where $\int d V$ is the volume integral. 

\section{Spectral expansion of the GK turbulence model}
\label{sec:spectralgenex}

This section presents the velocity-space spectral expansion used in \verb|GENE-X| to solve the edge and SOL GK turbulence model introduced in the previous section. We first describe the spectral expansion that we use in Section \ref{subsec:velocityspace}. Then, we derive the spectral formulation of the GK Vlasov-Maxwell equations in Section \ref{subsec:spectralvlasov} and \ref{subsec:spectralmaxwell}, respectively. The spectral expansion of the LBD collision operator is presented in Section \ref{subsec:spectrallbd}, while the fluid moments of the distribution function are obtained in Section \ref{subsec:momentsdistribution}. Finally, we discuss the energy consistency of the spectral approach in Section \ref{subsec:energyconservation}.

\subsection{Spectral velocity-space approach}
\label{subsec:velocityspace}

The spectral expansion utilizes a set of orthogonal Hermite and Laguerre polynomials, which are appropriately scaled in velocity-space \cite{frei2023,frei2023b}. In contrast to the grid approach of \verb|GENE-X|, which solves the GK Vlasov equation in the advection form, the spectral formulation presented in this work is based on the conservative form. More precisely, since the distribution function $f_\alpha$ and $B_\parallel^*$ appear together in Eq. (\ref{eq:vlasov}), we define $\mathcal{F}_\alpha = B_\parallel^* f_\alpha / B$. We then expand $\mathcal{F}_\alpha$ using scaled and normalized Hermite polynomials, denoted by $\hat H_p(\hat v_{\parallel \alpha}) = H_p(\hat v_{\parallel \alpha}) / \sqrt{2^p p!}$, and scaled Laguerre polynomials, denoted by $L_j(\hat \mu_\alpha)$, for the parallel and perpendicular velocity-space directions, respectively. The definitions and properties of the Hermite and Laguerre polynomials are summarized in \ref{appendix:properties}. Here, the normalized and scaled velocity-space coordinates $(\hat v_{\parallel \alpha}, \hat \mu_\alpha)$ are defined by $\hat v_{\| \alpha} = v_{\parallel} / v_{\tau_\alpha}$ and $\hat \mu_\alpha = \mu B / \tau_{\alpha}$, where the scaled thermal velocity $v_{\tau_\alpha} = \sqrt{2 \tau_\alpha / m_\alpha}$ depends on the scaled temperature $\tau_\alpha$. The scaled temperature $\tau_\alpha$ is a free parameter and enables us to scale the Hermite and Laguerre polynomials to properly resolve the temperature difference across the edge and SOL. In contrast to previous full-$f$ Hermite-Laguerre models (see, e.g., Ref. \cite{frei2020}), we assume $\tau_\alpha$ to be constant in both time and space, though it may differ between species. The convergence properties of the spectral expansion carried out in this section are discussed in more detail in \ref{appendix:properties}.

Using the scaled Hermite and Laguerre polynomial basis, we expand $\mathcal{F}_\alpha$ as

\begin{align} \label{eq:faexpansion}
  \mathcal{F}_\alpha = \sum_{p =0}^{\infty} \sum_{j =0}^\infty \mathcal{N}_\alpha^{pj} \hat H_p(\hat v_{\parallel \alpha}) L_j(\hat \mu_\alpha) F_{\mathcal{M} \alpha},
\end{align}
\\
where we introduce the scaled Maxwellian distribution function

\begin{align} \label{eq:fma}
 F_{\mathcal{M} \alpha}=   \frac{e^{- \hat v_{\parallel  \alpha }^2 - \hat \mu_\alpha}}{\pi^{3/2} v_{\tau_\alpha}^{3}}.
\end{align}
\\
In Eq. (\ref{eq:faexpansion}), we introduce the spectral coefficients, $\mathcal{N}_\alpha^{pj}$, which can be expressed by using the orthogonality relations of the Hermite and Laguerre polynomials, Eqs. (\ref{eq:orthohermite}) and (\ref{eq:ortholaguerre}). It yields

\begin{align} \label{eq:npj}
  \mathcal{N}_\alpha^{pj}  = \int d W  \hat H_p(\hat v_{\parallel \alpha}) L_j(\hat \mu_\alpha) f_\alpha.
\end{align}
\\
We now aim to find an evolution equation of the spectral coefficients $\mathcal{N}_\alpha^{pj}$.

\subsection{Spectral GK Vlasov equation}
\label{subsec:spectralvlasov}

 To derive an evolution equation for $\mathcal{N}_\alpha^{pj}$, we first observe that the equations of motion, given in Eqs. (\ref{eq:motions}), contain terms that are divided by $B_\parallel^*$. When projected onto the scaled Hermite-Laguerre basis, these terms implicitly yield coupling between all spectral coefficients because of divisions between the velocity-space coordinates $(v_\parallel, \mu)$. Hence, to avoid such terms divisions, we use the expansion 

\begin{align} \label{eq:bstarapprox}
  \frac{\bm B^*}{B_\parallel^*} \simeq \frac{\bm B}{B} + m_\alpha v_{\parallel} \frac{c}{q_\alpha B} \grad \times \bm b + \frac{1}{B}\grad A_{1 \parallel} \times \bm b,
\end{align}
\\
in Eqs. (\ref{eq:motions}) where we approximate $1/B_{\parallel}^* \simeq 1/B$, neglecting the guiding-center correction term in the denominator. We remark that the approximation in Eq. (\ref{eq:bstarapprox}) does not contradict the energy consistency of the GK Vlasov-Maxwell system, as discussed in Section \ref{subsec:energyconservation}. We now use Eq. (\ref{eq:bstarapprox}) and multiply the GK Vlasov equation in the conservative form, Eq. (\ref{eq:vlasov}), by the scaled Hermite-Laguerre basis, $\hat H_p( \hat v_{\parallel \alpha} )  L_j(\hat \mu_\alpha )$, and integrate over velocity-space. When performing the velocity-space integrals, the gradient operator commutes with $\hat H_p( \hat v_{\parallel \alpha} )$ but not with $L_j(\hat \mu_\alpha )$. This is due to the dependence of $\hat \mu_\alpha$ on the magnetic field strength $B$. The result is

\begin{align} \label{eq:spectralvlasov}
  \frac{\partial}{\partial t} \mathcal{N}_\alpha^{pj} & + \grad \cdot \left(  \bm \Gamma_{\alpha \parallel}^{pj} +\bm \Gamma_{\alpha \kappa}^{pj} + \bm \Gamma_{\alpha \grad}^{pj} + \bm \Gamma_{\alpha 1}^{pj}\right) \nonumber \\
& + \sqrt{\frac{ \tau_\alpha}{m_\alpha}}   \frac{1}{B} \grad_{ 1 \parallel  }B   \left[  \sqrt{2} j \left(   \mathbb{I}^{pj-1}_{1\ell k} - \mathbb{I}^{pj}_{1\ell k}  \right) + \sqrt{p} \ \mathbb{I}^{p-1j}_{\ell k1}  \right]\mathcal{N}_\alpha^{\ell k} \nonumber \\
& +  \frac{ c \tau_\alpha }{q_\alpha B^2} \grad \times \bm b \cdot \grad B   \left[2 j \left(  \mathbb{I}^{pj-1 }_{2\ell k} -  \mathbb{I}^{pj }_{2\ell k} \right) + \sqrt{2p} \ \mathbb{I}^{p-1j}_{1\ell k1}\right] \mathcal{N}_\alpha^{\ell k}  \nonumber \\
&  +  \frac{c}{ B^2 } \bm b \times \grad \psi_{1 \alpha} \cdot \grad B  j \left(  \mathbb{I}^{pj -1}_{\ell k} - \mathbb{I}^{pj }_{\ell k} \right)  \mathcal{N}_\alpha^{\ell k} \nonumber \\
&  +  \frac{ c}{ B} \grad \times \bm b \cdot \grad \psi_{1 \alpha} \sqrt{2p} \ \mathbb{I}^{p-1j }_{1\ell k} \mathcal{N}_\alpha^{\ell k}  \nonumber \\
&  + \frac{q_\alpha}{\sqrt{m_\alpha \tau_\alpha}}  \grad_{ 1 \parallel  } \psi_{1 \alpha} \sqrt{p} \ \mathbb{I}^{p-1j }_{\ell k} \mathcal{N}_\alpha^{\ell k}  \nonumber \\
& + \frac{q_\alpha \sqrt{p}}{c \sqrt{m_\alpha \tau_\alpha}  } \frac{\partial A_{1 \parallel}}{\partial t} \  \mathbb{I}^{p-1j }_{\ell k} \mathcal{N}_\alpha^{\ell k} = \sum_\beta C_{\alpha \beta }^{pj},
\end{align}
\\
with generalized fluxes defined by

\begin{subequations} \label{eq:generalizedfluxes}
\begin{align}
    \bm \Gamma_{\alpha \parallel}^{pj} &  = \sqrt{\frac{2 \tau_\alpha}{m_\alpha}}  \frac{\bm B_{T}}{B} \ \mathbb{I}^{pj}_{1\ell k} \mathcal{N}_\alpha^{\ell k}, \label{eq:parallelflux}  \\
    \bm \Gamma_{\alpha \kappa}^{pj} & = \frac{2 \tau_\alpha  c}{q_\alpha B} \grad \times \bm b  \ \mathbb{I}^{pj }_{2\ell k} \mathcal{N}_\alpha^{\ell k}, \label{eq:curvatureflux} \\
    \bm \Gamma_{\alpha \grad}^{pj} & = \frac{c \tau_\alpha }{q_\alpha } \frac{\bm b}{B^2} \times \grad B \ \mathbb{I}^{pj}_{\ell k1} \mathcal{N}_\alpha^{\ell k}, \label{eq:gradbflux} \\
    \bm \Gamma_{\alpha 1}^{pj} & = \frac{c}{ B } \bm b \times \grad \psi_{1 \alpha}\  \mathbb{I}^{pj }_{\ell k} \mathcal{N}_\alpha^{\ell k}. \label{eq:exbflux}
\end{align}
\end{subequations}
\\
In Eqs. (\ref{eq:spectralvlasov}) and (\ref{eq:generalizedfluxes}), the Einstein summation convention over $(\ell, k)$ indices is assumed. In Eq. (\ref{eq:spectralvlasov}), the total parallel gradient operator is denoted by $\grad_{1 \parallel} = (\bm B_{T} / B) \cdot \grad = \bm b \cdot \grad + (\grad A_{1 \parallel} \times \bm b) / B \cdot \grad$, which includes the contribution from magnetic flutter. The expressions for the parallel gradient operator and the differential operators used to compute the divergence of the generalized fluxes are provided in \ref{appendix:differentialoperators}. These are formulated in terms of Poisson brackets and curvature operators, which are discretized in \verb|GENE-X| using locally field-aligned coordinates.

We note that the generalized fluxes, given in Eq. (\ref{eq:generalizedfluxes}), represent the spectral formulations of the velocity-space fluxes associated with parallel motion along the magnetic field lines in Eq. (\ref{eq:parallelflux}), curvature drift in Eq. (\ref{eq:curvatureflux}), magnetic gradient drift in Eq. (\ref{eq:gradbflux}), and the generalized drift driven by the potential $\psi_{1 \alpha}$ in Eq. (\ref{eq:exbflux}). The remaining terms arise from parallel acceleration and magnetic pumping effects \cite{scott2021}, which result from the spectral projection of the term $\hat \mu_\alpha \dot{\bm{R}} \cdot \grad B$, related to the gradient of $L_j(\hat \mu_\alpha)$.

The coupling between the coefficients in the spectral GK Vlasov equation, Eq. (\ref{eq:spectralvlasov}), and in the generalized fluxes, Eq. (\ref{eq:generalizedfluxes}), are expressed through the terms, $\mathbb{I}^{ pj }_{\beta\ell k \gamma} \mathcal{N}_\alpha^{\ell k}$, defined by

\begin{align}
 \mathbb{I}^{ pj }_{\beta\ell k \gamma} \mathcal{N}_\alpha^{\ell k} = \int d W f_\alpha  \hat v_{\parallel \alpha}^\beta \hat{H}_p(\hat v_{\parallel \alpha}) \hat \mu_\alpha^\gamma  L_j(\hat \mu_\alpha),
\end{align}
\\
where $\beta$, $\gamma$ are positive integers. While an analytical expression of $\mathbb{I}^{ pj }_{\beta\ell k \gamma} $ can be derived for arbitrary $\beta$ and $\gamma$, we focus here on deriving closed expressions for the ones that explicitly appear in Eqs. (\ref{eq:spectralvlasov}) and (\ref{eq:generalizedfluxes}). Using the recurrence properties of the Hermite and Laguerre polynomials \cite{gradshteyn2014}, we obtain 

\begin{subequations} \label{eq:mixingmatrices}
  \begin{align}
    \mathbb{I}^{pj }_{\ell k}& = \delta_\ell^p \delta_{k}^j, \\
    \mathbb{I}^{pj }_{1\ell k}& = \sqrt{\frac{(p+1)}{2}} \delta_\ell^{p+1}  \delta_{k}^j+  \sqrt{\frac{p}{2}}  \delta_\ell^{p-1}  \delta_{k}^j, \label{eq:Ipj1lk}\\
      \mathbb{I}^{pj }_{2\ell k}
  &  =  \frac{1}{2}  \sqrt{(p+2)(p+1)} \delta_\ell^{p+2} \delta_{k}^j \nonumber \\& + \left( p + \frac{1}{2}\right) \delta_\ell^{p}\delta_{k}^j  + \frac{1}{2} \sqrt{p(p-1)} \delta_\ell^{p-2} \delta_{k}^j, \label{eq:Ipj2lk}\\
      \mathbb{I}^{pj }_{\ell k1}  & = (2 j+1) \delta_\ell^{p}\delta_{k}^{j} -j \delta_\ell^{p} \delta_{k}^{j-1}- (j+1)\delta_\ell^{p}\delta_{k}^{j+1}, \label{eq:Ipjlk1} \\
      \mathbb{I}^{pj }_{1\ell k1}  & = (2 j + 1)  \mathbb{I}^{pj }_{1\ell k}  -j  \mathbb{I}^{ pj-1 }_{1\ell k} - (j+1)  \mathbb{I}^{pj+1 }_{1\ell k}\label{eq:Ipjl1k1}, \end{align}
\end{subequations}
\\
where the indices $\beta$ and $\gamma$ are omitted when equal to $0$ and the Kronecker delta $\delta_i^j$ is used. We remark that Eq. (\ref{eq:Ipj1lk}) represents the coupling due to parallel streaming (proportional to $v_\parallel$) along the magnetic field lines, while $\mathbb{I}^{pj }_{2\ell k}$ in Eq. (\ref{eq:Ipj2lk}) is associated with curvature drift (proportional to $v_\parallel^2$). Similarly, the dynamics of trapped particles, driven by the magnetic mirror force, result in the coupling described by $\mathbb{I}^{pj }_{\ell k1}$ in Eq. (\ref{eq:Ipjlk1}).

\subsection{Spectral GK Maxwell equations}
\label{subsec:spectralmaxwell}

We now derive the spectral formulation of the GK Maxwell equations given in Eqs. (\ref{eqs:fields}) and (\ref{eq:ohm}). Using the definition of the spectral coefficients, $\mathcal{N}_\alpha^{pj}$ in Eq. (\ref{eq:npj}), the velocity-space integrals in the GK Maxwell equations can be evaluated analytically. Thus, the spectral formulation of the QN condition, Ampere’s law, and Ohm’s law are

\begin{subequations} \label{eq:spectralmaxwells}
  \begin{align}
 -\boldsymbol{\nabla} \cdot\left( \sum_\alpha   \frac{  m_\alpha  c^2 \mathcal{N}_\alpha^{00}}{B^2}   \nabla_{\perp}  \phi_1\right) & =\sum_\alpha   q_\alpha \mathcal{N}_\alpha^{00}   \label{eq:poissonvspec} ,\\
 -  \Delta_{\perp}  A_{1 \|} & =   4 \pi  \sum_\alpha  \frac{q_\alpha}{c}   \sqrt{ \frac{\tau_{\alpha}}{  m_\alpha}}    \mathcal{N}_\alpha^{10}  \label{eq:amperevspec} , \\
   - \left(   \frac{\Delta_{\perp} }{4 \pi} -   \sum_\alpha \frac{q_\alpha^2 \mathcal{N}_\alpha^{00}}{c^2 m_\alpha}   \right)\frac{\partial A_{1 \parallel} }{\partial t}  & =    \sum_\alpha \frac{q_\alpha}{c}   \sqrt{ \frac{\tau_{\alpha}}{  m_\alpha}}  \left( \frac{\partial \mathcal{N}_\alpha^{10}}{\partial t}\right)^* \label{eq:ohmvspec},
 \end{align}
\end{subequations}
\\
respectively. We note that, in Ohm's law Eq. (\ref{eq:ohmvspec}), the notation $(\partial_t \mathcal{N}_\alpha^{10})^*$ represents the static part of the spectral GK Vlasov equation, which is obtained by setting $(p,j) = (1,0)$ in Eq. (\ref{eq:spectralvlasov}).

\subsection{Spectral LBD collision operator model}
\label{subsec:spectrallbd}

In this work, we focus on the spectral expansion of the LBD collision operator only. One advantage of this collision operator model compared to the nonlinear Landau collision operator \cite{landau1936} is that the former has a straightforward representation in the Hermite-Laguerre polynomial basis while the latter is more involved (see, e.g., Refs. \cite{jorge2019,frei2021d}). Indeed, the spectral representation of the LBD collision operator \cite{ulbl2022} reads

\begin{align} \label{eq:lbdvspec}
 \mathcal{C}_{\alpha \beta}^{pj}   =   \nu_{\alpha \beta} &\left[   -   \left(   p   + 2  j   \right)  \mathcal{N}_\alpha^{pj}  \nonumber \right. \nonumber \\ & \left.   + \left( \bar  T_{\alpha \beta} - 1 \right)  \left( \sqrt{p (p-1)}    \mathcal{N}_\alpha^{p-2 j }     - 2 j  \mathcal{N}_\alpha^{p j-1 }  \right)  \nonumber \right. \nonumber \\ & \left. + \bar u_{\alpha \beta}    \sqrt{2 p}  \mathcal{N}_\alpha^{p-1 j}  \right].
\end{align}
\\
In Eq. (\ref{eq:lbdvspec}), the mixing temperature, $ \bar  T_{\alpha \beta}$, and flow velocity, $\bar u_{\alpha \beta}$, variables are given by \cite{ulbl2022}

\begin{subequations} \label{eq:mixingvars}
\begin{align}
\bar u_{\alpha \beta} & = \frac{1}{ m_\alpha  + m_\beta} \left(  m_\alpha \bar u_\alpha+ \sqrt{   m_\beta  m_\alpha \frac{ \tau_{  \beta }}{  \tau_{ \alpha}}} \bar u_\beta \right) , \\
\bar T_{\alpha \beta} & = \frac{1}{2} \left(\bar T_\alpha + \frac{ \tau_{\beta}}{\tau_{\alpha} } \bar T_\beta \right)-\frac{1}{3} \left[ \left( 1 + \frac{ m_\beta}{ m_\alpha}  \right) \bar u_{\alpha \beta}^2-\bar u_\alpha^2 -  \frac{ \tau_{ \beta}}{ \tau_{\alpha}} \bar u_\beta^2 \right],
\end{align}
\end{subequations}
\\
with single species flow velocities and temperatures

\begin{subequations} \label{eq:mixingvariable}
\begin{align}
\bar u_{\alpha} = \frac{ \mathcal{N}_\alpha^{10} }{ \sqrt{2} \mathcal{N}_\alpha^{00}},
\end{align}
\begin{align}
\bar T_\alpha =  \frac{1}{3 \mathcal{N}_\alpha^{00}} \left( \sqrt{2} \mathcal{N}_\alpha^{20} + 3 \mathcal{N}_\alpha^{00} - 2 \mathcal{N}_\alpha^{01}\right) - \frac{1}{3}  \left( \frac{ \mathcal{N}_\alpha^{10} }{  \mathcal{N}_\alpha^{00}}\right)^2.
\end{align}
\end{subequations}
\\
The conservation of single species particles, total momentum, and total energy of the collision operator $\mathcal{C}_{\alpha \beta}^{pj}$ reads

\begin{subequations} \label{eq:collisions}
  \begin{align}
    \mathcal{C}^{00}_{\alpha \beta} & = 0, \\
    \sqrt{m_\alpha \tau_\alpha}  \mathcal{C}^{10}_{\alpha \beta} & = -   \sqrt{m_\beta \tau_\beta}  \mathcal{C}^{10}_{\beta \alpha}, \\
\tau_\alpha \left(\frac{\mathcal{C}_{\alpha \beta}^{20}}{\sqrt{2}} - \mathcal{C}_{\alpha \beta}^{01}  \right)  & = - \tau_\beta \left( \frac{\mathcal{C}_{\beta \alpha}^{20}}{\sqrt{2}} - \mathcal{C}_{\beta \alpha}^{01} \right) . \label{eq:collenergyconservation}
  \end{align}
\end{subequations}
\\
With the analytical expressions of the mixing variables in Eq. (\ref{eq:mixingvars}), it can be shown that the conservation laws given in Eq. (\ref{eq:collisions}) are exactly satisfied by the spectral LBD collision operator.

\subsection{Fluid moments of the full-$f$ distribution function}
\label{subsec:momentsdistribution}

We now express the fluid moments of the full-$f$ distribution function $f_\alpha$ in terms of the spectral coefficients $\mathcal{N}_\alpha^{pj}$. More precisely, we consider the gyrocenter density $n_\alpha$, parallel flow velocity $u_{\parallel \alpha}$, parallel and perpendicular (lab frame) temperatures, $T_{\parallel \alpha}$ and $T_{\perp \alpha}$, as well as the parallel heat fluxes of parallel and perpendicular energy, $Q_{\parallel \parallel \alpha}$ and $Q_{\parallel \perp \alpha}$, respectively. Using Eq. (\ref{eq:npj}), we derive

\begin{subequations} \label{eq:mom2da}
\begin{align}
 n_\alpha  & = \int d W  f_\alpha  = \mathcal{N}_\alpha^{00}, \\
n_\alpha u_{\parallel \alpha } &  =  \int d W v_{\| } f_\alpha  =  \sqrt{ \frac{ \tau_{\alpha} }{m_\alpha } }   \mathcal{N}_\alpha^{10}, \label{eq:upara} \\
n_\alpha  T_{\parallel \alpha} & =   \int d W   m_\alpha \left(  v_{\|}-u_{\parallel \alpha}  \right)^2 f_\alpha \nonumber \\
& =   \tau_\alpha  \left(  \sqrt{2} \mathcal{N}_\alpha^{20} +  n_\alpha \right)  -  n_\alpha  m_\alpha  u_{\parallel \alpha } ^2,   \\
n_\alpha T_{\perp \alpha} & = \int d W f_\alpha \mu B  =  \tau_{\alpha}   \left(n_\alpha - \mathcal{N}_\alpha^{01} \right), \\
Q_{\parallel \parallel \alpha} & =    \int d W  \frac{m_\alpha}{2}  v_\parallel^3 f_\alpha  \nonumber \\
& =  \sqrt{\frac{ \tau_\alpha^3}{ m_\alpha} }   \left(\sqrt{\frac{3}{2}} \mathcal{N}_\alpha^{30}  + \frac{3}{2} \mathcal{N}_\alpha^{10} \right) , \\
Q_{\parallel \perp \alpha} & =  \int d W   v_\parallel \mu B  f_\alpha  \nonumber \nonumber \\
& =    \sqrt{\frac{ \tau_{\alpha}^3}{m_\alpha} } \left(  \mathcal{N}_\alpha^{10} - \mathcal{N}_\alpha^{11}\right).
\end{align}
\end{subequations}
\\
We remark that the total temperature, $T_\alpha$, is then defined by $ T_{\alpha} = ( T_{\parallel \alpha}  + 2 T_{\perp \alpha})/3 $. From the expressions in Eqs. (\ref{eq:mom2da}), it can be observed that densities and temperatures only involve the spectral coefficients with even power in velocity variables. On the other hand, the parallel velocities and fluxes depend on coefficients with odd powers in velocity variables. Hence, the spectral coefficients, which are even power ($p$ even), are referred to as state variables, while the coefficients with odd power ($p$ odd) are referred to as flux variables \cite{scott2021}.

\subsection{Spectral energy conservation}
\label{subsec:energyconservation}

 Using the definitions of the fluid moments in Eq. (\ref{eq:mom2da}), the total energy $\mathcal{E}$ of the GK Vlasov-Maxwell system, given in Eq. (\ref{eq:totalenergy}), can be expressed as

\begin{align} \label{eq:spectralenergy}
  \mathcal{E} & = \sum_\alpha \int dV \left[ \frac{\tau_\alpha}{2} \left( \sqrt{2} \mathcal{N}_\alpha^{20} + 3 \mathcal{N}_\alpha^{00} - 2 \mathcal{N}_\alpha^{01} \right) + q_\alpha n_\alpha \psi_{1 \alpha} \right] \nonumber \\
  & + \int dV \frac{1}{8 \pi} \left|\nabla_{\perp} A_{1 \|}\right|^2.
\end{align}
\\
The spectral formulation used in this work preserves the conservation of total energy, $\mathcal{E}$ exactly. In fact, this can be demonstrated by explicitly evaluating the time derivative of Eq. (\ref{eq:spectralenergy}) and using the spectral GK Vlasov equation, Eq. (\ref{eq:spectralvlasov}) to obtain explicit expressions of the time evolution of $\mathcal{N}_\alpha^{20}$, $\mathcal{N}_\alpha^{01}$, and $\mathcal{N}_\alpha^{00}$. The collisional terms cancel out due to the exact energy conservation of the LBD collision operator (see Eq. (\ref{eq:collenergyconservation})). After spatial integration, contributions from the generalized fluxes that are given in Eqs. (\ref{eq:generalizedfluxes}) and appear under the divergence operator vanish. The remaining terms combine and cancel exactly, illustrating that the magnetic pumping terms effectively redistribute energy among the spectral coefficients while conserving the total energy. Finally, the contribution from the induction term $\partial_t A_{1 \parallel}$, the time derivative of the last term in Eq. (\ref{eq:spectralenergy}), is canceled by the dynamical part of the $\mathcal{N}_\alpha^{20}$ equation.

\section{Numerical aspects and implementation}
\label{sec:numericalimplementation}

This section describes the numerical aspects and implementation of the spectral approach in the \verb|GENE-X| code. We start with an overview of the discretization scheme used in the grid approach of \verb|GENE-X| in Section \ref{subsec:discretization}. Then, we discuss the code design in Section \ref{subsec:codedesign}. The numerical hyperdiffusion and dissipation are detailed in Section \ref{subsec:hyperdiffusionandissipation}, while the time stepping and boundary conditions are discussed in Sections \ref{subsec:timestepping} and \ref{subsec:boundaryconditions}, respectively. The normalization used in \verb|GENE-X| is outlined in \ref{appendix:normalization}.

\subsection{Discretization schemes}
\label{subsec:discretization}

We introduce the velocity-space and configuration discretization schemes used in the grid approach of \verb|GENE-X|. While more details can be found in Ref. \cite{michels2021}, we report only the most important aspects to ease the comparison with the spectral approach. 

The velocity-space is discretized on a fixed grid using the coordinates $v_\parallel$ and $\mu$ in the grid approach. The coordinates $(v_\parallel, \mu)$ are normalized such that $\hat v_\parallel = v_\parallel \sqrt{2 \Tref / m_\alpha}$ and $\hat \mu = \mu \Bref / \Tref$, where $\Tref$ is the constant reference temperature, $\Bref$ is the reference magnetic field strength (see \ref{appendix:normalization} for the normalization). An equidistant and cartesian grid is then constructed with grid points, denoted by $(\hat v_\parallel^{(\ell)}, \hat \mu^{(m)})$ and labeled with the indices $\ell = 1, 2, \dots, N_{v_\parallel}$ and $m = 1, 2, \dots, N_\mu$. The distribution function $f_\alpha$ is therefore calculated at the velocity space grid points $(\hat v_\parallel^{(\ell)}, \hat \mu^{(m)})$ (see Fig. \ref{fig:grid}). Here, $N_{v_\parallel}$ and $N_\mu$ are the total number of velocity-space grid points in the $v_\parallel$ and $\mu$ directions, respectively. We note that, in this work, $N_{v_\parallel}$ and $N_\mu$ will refer to both the number of grid points and the number of spectral coefficients in the case of the spectral approach. A fourth-order centered finite difference scheme is used to numerically compute the first-order derivative in $v_\parallel$ that appears in the collisionless part of the GK Vlasov equation. On the other hand, the computation of the collisional part, which includes second-order velocity-space derivatives and integrals of $f_\alpha$, is obtained by a second-order finite volume scheme \cite{ulbl2022} (see Fig. \ref{fig:grid}). Finally, Simpson and Gauss-Laguerre quadrature schemes are used in the $\hat v_\parallel$ and $\hat \mu$ directions, respectively, to compute the fluid moments of $f_\alpha$ required for the GK Maxwell equations.

The discretization of the configuration space is performed using the FCI approach \cite{hariri2013,stegmeir2016}, which employs a locally field-aligned coordinate system \cite{michels2021}. In this system, one coordinate $y$ is aligned with the magnetic field, while the other two $R$ and $Z$ (flux coordinate independent) describe the perpendicular direction in the poloidal planes. The grid-spacing in the poloidal planes is thus denoted by $\Delta RZ$, while the number of poloidal planes is $N_{\varphi}$. Parallel derivatives are constructed by tracing magnetic field lines between poloidal planes and performing interpolation within each plane. Then, a fourth-order centered finite difference scheme is used. To discretize the differential operators appearing in the spectral GK Vlasov equation, Eq. (\ref{eq:spectralvlasov}), (detailed in \ref{appendix:differentialoperators}) and other perpendicular derivatives, a fourth-order centered finite difference scheme is employed, while a second-order Arakawa scheme \cite{arakawa1997} handles the nonlinear advection terms associated with the $\bm E \times \bm B$ drift and flutter terms. Finally, the GK Maxwell equations are solved using an elliptic solver that employs the multigrid algorithm \cite{stegmeir2023,ulbl2023}. We remark that the same configuration space discretization scheme as in the grid approach is used in the spectral implementation. 

\subsection{Code design}
\label{subsec:codedesign}

The object-oriented design of \verb|GENE-X| offers the flexibility for integrating the spectral approach with minimal code rewrites. The numerical evaluations of the right-hand side of the GK Vlasov equation, moment calculations, and collisions, are represented as operators encapsulated in abstract classes that act on the distribution functions and electromagnetic fields \cite{michelsphd2021}. Grid and spectral implementations are then defined in derived classes, enabling efficient testing and comparison of different numerical formulations.

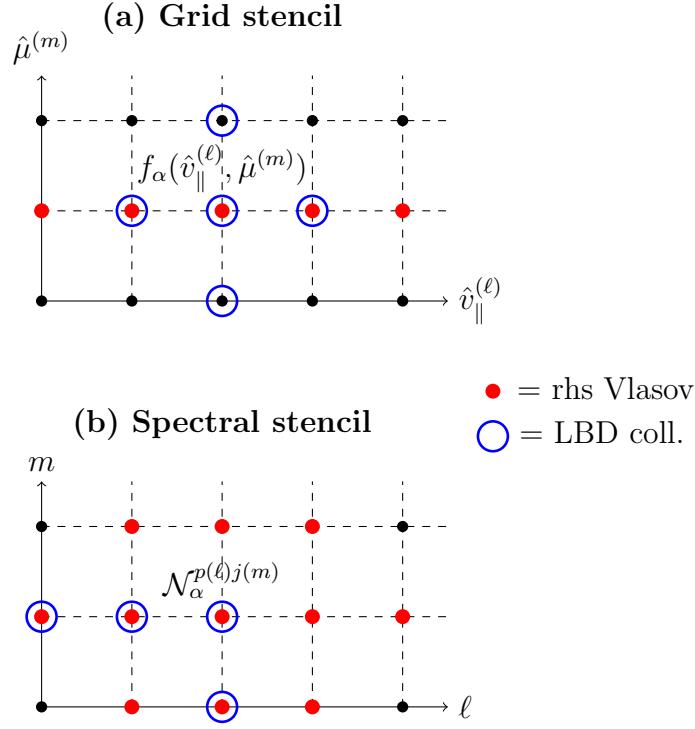
\begin{figure}[h]
\centering
  \begin{tikzpicture}[scale = 1.2]
  	  \def\XShift{0}
      \def\YShift{-4.5}
      % Draw axes
      \draw[->] (0,0) -- (4.5,0) node[right] {$\hat v_\parallel^{(\ell)}$};
      \draw[->] (0,0) -- (0,2.5) node[above] {$\hat \mu^{(m)}$};
      \draw[dashed] (0,1) -- (4.5,1) node[right] {};
      \draw[dashed] (0,2) -- (4.5,2) node[right] {};
      \draw[dashed] (1,0) -- (1, 2.5) node[above] {};
      \draw[dashed] (2,0) -- (2, 2.5) node[above] {};
      \draw[dashed] (3,0) -- (3, 2.5) node[above] {};
      \draw[dashed] (4,0) -- (4, 2.5) node[above] {};
    
      % Label grid points
          \foreach \x in {0,1,2,3,4}
          \foreach \y in {0,1,2}
              \node[fill=black, circle,inner sep = 1.5pt] at (\x,\y) {};

      % Label the compute node
      \node[fill=red, circle, inner sep=2pt,label={$f_\alpha(\hat v_\parallel^{(\ell)}, \hat \mu^{(m)}$)}] at (2,1) {};

      % Grid ghost points
      \node[fill=red, circle, inner sep=2pt, label={}] at (0,1) {};
      \node[fill=red, circle, inner sep=2pt, label={}] at (1,1) {};
      \node[fill=red, circle, inner sep=2pt, label={}] at (3,1) {};
      \node[fill=red, circle, inner sep=2pt, label={}] at (4,1) {};
      
      \node[fill=none, draw=blue, circle, line width = 1pt,inner sep=4pt, label={}] at (2,2) {};
        \node[fill=none, draw=blue, circle, line width = 1pt,inner sep=4pt, label={}] at (2,0) {};
        \node[fill=none, draw=blue, circle, line width = 1pt,inner sep=4pt, label={}] at (2,1) {};
     \node[fill=none, draw=blue, circle, line width = 1pt,inner sep=4pt, label={}] at (3,1) {};
      \node[fill=none, draw=blue, circle, line width = 1pt,inner sep=4pt, label={}] at (1,1) {};

\node[fill=red, circle, inner sep=2pt, label={[xshift=1.5cm, yshift=-0.35cm]$=$ rhs Vlasov}] at (5,-1) {};
\node[fill=none, circle, inner sep=4pt, draw=blue, label={[xshift=1.45cm, yshift=-0.45cm]$=$ LBD coll.}, line width =1pt] at (5,-1.5) {};

      \node[inner sep=3pt,label={\textbf{(a) Grid stencil}}] at (2,2.75) {};

      % Draw axes
      \draw[->] (\XShift,\YShift) -- (\XShift + 4.5,\YShift) node[right] {$\ell$};
      \draw[->] (\XShift,\YShift) -- (\XShift,\YShift + 2.5) node[above] {$m$};
      \draw[dashed] (\XShift,\YShift+1) -- (\XShift +4.5,\YShift+1) node[right] {};
      \draw[dashed] (\XShift,\YShift+2) -- (\XShift +4.5,\YShift+2) node[right] {};
      \draw[dashed] (\XShift + 1,\YShift) -- (\XShift +1, \YShift+2.5) node[above] {};
      \draw[dashed] (\XShift +2,\YShift) -- (\XShift +2, \YShift+2.5) node[above] {};
      \draw[dashed] (\XShift +3,\YShift) -- (\XShift +3, \YShift+2.5) node[above] {};
      \draw[dashed] (\XShift +4,\YShift) -- (\XShift +4, \YShift+2.5) node[above] {};
            % Label grid points
          \foreach \x in {0,1,2,3,4}
          \foreach \y in {0,1,2}
              \node[fill=black, circle, inner sep = 1.5pt] at (\XShift + \x,\YShift +\y) {};

      % Label the compute node
      \node[fill=red, circle,inner sep=2pt, label={$\mathcal{N}_\alpha^{p(\ell) j(m)}$}] at (\XShift + 2,\YShift +1) {};

      % Grid ghost points
      \node[fill=red, circle, inner sep=2pt, label={}] at (\XShift,\YShift +1) {};
      \node[fill=red, circle, inner sep=2pt, label={}] at (\XShift +1,\YShift +1) {};
      \node[fill=red, circle, inner sep=2pt, label={}] at (\XShift +3,\YShift +1) {};
      \node[fill=red, circle, inner sep=2pt, label={}] at (\XShift +4,\YShift +1) {};
      \node[fill=red, circle, inner sep=2pt, label={}] at (\XShift +2,\YShift +0) {};
      \node[fill=red, circle, inner sep=2pt, label={}] at (\XShift +2,\YShift +2) {};
      \node[fill=red, circle, inner sep=2pt, label={}] at (\XShift +3,\YShift +2) {};
      \node[fill=red, circle, inner sep=2pt, label={}] at (\XShift +3,\YShift +0) {};
      \node[fill=red, circle, inner sep=2pt, label={}] at (\XShift +1,\YShift +2) {};
      \node[fill=red, circle, inner sep=2pt, label={}] at (\XShift +1,\YShift +0) {};

    \node[fill=none, circle, draw=blue, inner sep=4pt, label={}, line width =1pt] at (\XShift +2,\YShift +1) {};
    \node[fill=none, circle, draw=blue, inner sep=4pt, label={}, line width =1pt] at (\XShift +1,\YShift +1) {};
    \node[fill=none, circle, draw=blue, inner sep=4pt, label={}, line width =1pt] at (\XShift,\YShift +1) {};
    \node[fill=none, circle, draw=blue, inner sep=4pt, label={}, line width =1pt] at (\XShift + 2,\YShift) {};
    
      \node[inner sep=3pt,label={\textbf{(b) Spectral stencil}}] at (\XShift + 2,\YShift + 2.75) {};
  \end{tikzpicture}
  \caption{Velocity-space stencils associated with the grid (a) and spectral (b) implementations. The red circles represent the points used for the computation of the right-hand side (rhs) of the collisionless part of the GK Vlasov equation, while the empty blue circles denote the points used to calculate the LBD collision operator. In the grid approach, a second-order finite volume scheme is used for collisions (see Ref. \cite{ulbl2022}) while Eq. (\ref{eq:lbdvspec}) is utilized for the spectral approach. }
  \label{fig:grid}
\end{figure}

The data structures and algorithms used in the grid implementation require minimal adjustments for the spectral formulation. The electromagnetic field structure is unchanged, and the five-dimensional array for the distribution function $f_\alpha$ at a velocity-space grid point $(\hat{v}_\parallel^{(\ell)}, \hat{\mu}^{(m)})$ can be mapped directly to spectral coefficients $\mathcal{N}_\alpha^{p(\ell) j(m)}$. Here, $\ell = 1, \dots, N_{v_\parallel}$ and $m = 1, \dots, N_\mu$ denote indices in the velocity-space discretization. This approach eliminates redundant data structures, reusing grid algorithms efficiently for the spectral method. Loops over velocity-space grid points adapt to loops over $\mathcal{N}_\alpha^{pj}$, enabling a unified implementation. Additionally, the grid stencil, designed for finite differences in velocity-space derivatives, can be adapted for spectral formulation due to the similar sizes and ghost-point arrangements of both stencils. This includes two-sided ghosts in $p$ for $\mathbb{I}_{2\ell k}^{pj}$ and one-sided ghosts in $j$ for $\mathbb{I}_{\ell k1}^{pj}$, along with edge ghosts for coupling between spectral coefficients in magnetic pumping terms, as illustrated in Fig. (\ref{fig:grid}).

%As a result, the spectral approach can reuse the same MPI decomposition and communication strategy for parallelization along the $(p, j)$ dimensions as in the $(\hat{v}_\parallel, \hat{\mu})$ dimensions.

\subsection{Hyperdiffusion and numerical dissipation}
\label{subsec:hyperdiffusionandissipation}

Similar to the grid approach, hyperdiffusion and dissipation in both the configuration and velocity-space are necessary to remove spurious numerical artifacts associated with finite resolution and to ensure stable numerical integration.

Since the spectral approach utilizes the same discrete operators with the FCI scheme in configuration space as the grid approach, a similar method is considered to stabilize high-frequency and grid-scale modes, as described in Ref. \cite{michels2021}. More precisely, numerical fourth order, bi-harmonic hyperdiffusion is introduced along the parallel direction to the magnetic field and in the poloidal plane on the right-hand side of the spectral GK Vlasov equation. In addition, a buffer region with second-order harmonic diffusion is also added near the boundary of the computational domain in the configuration space, ensuring a smooth transition between the prescribed boundary conditions (see Section (\ref{subsec:boundaryconditions})) and the computational domain \cite{michels2021}.

Although the spectral approach does not require numerical diffusion to stabilize the velocity-space finite difference scheme, some dissipation is nevertheless necessary to prevent the appearance of spurious numerical artifacts associated with the finite number of spectral coefficients used in the simulations. These artifacts include, for instance, recurrence effects \cite{frei2023} and energy accumulation in the highest-order spectral coefficients. Therefore, an artificial energy sink must be introduced. The simplest possible model introduces a diagonal damping term, $S_\alpha^{pj}$, in the spectral GK Vlasov equation, which mimics Landau damping and can be defined by \cite{scott2021}

\begin{align} \label{eq:landauclosure}
S_\alpha^{pj} = - \sqrt{\frac{2 \tau_\alpha}{m_\alpha}} K_\parallel \sqrt{p} \mathcal{N}^{pj},
\end{align}
\\
The term $S_\alpha^{pj}$ is added to the right-hand side of the spectral GK Vlasov equation for the flux variables ($p$ odd) with $p + 2j \geq N_{v_\parallel} - 1$ to ensure the conservation of the total energy $\varepsilon$ defined in Eq. (\ref{eq:totalenergy}). Here, $K_\parallel$ is a positive constant associated with the typical parallel scale length of the Landau damping, which is inversely proportional to the connection length. It should be noted that the dependency of the prefactor of Eq. (\ref{eq:landauclosure}) stems from the typical speed of the wave associated with the flux variable, with odd $p$, and its state variable, with even $p - 1$ \cite{scott2021}. Although more sophisticated models could be constructed, the damping term in Eq. (\ref{eq:landauclosure}) is sufficient for the present work. We finally remark that the value of $K_\parallel$ does not affect the saturated turbulent state, but helps to stabilize the initial phase of the simulations in particular at low $N_{v_\parallel}$. On the other hand, this damping term has negligible effects when increasing $N_{v_\parallel}$.

\subsection{Time stepping}
\label{subsec:timestepping}

To advance the spectral GK Vlasov-Maxwell system in time, an explicit fourth-order Runge-Kutta (RK4) scheme is used. While the  \verb|GENE-X| code offers a variety of time integration schemes, the stability provided by the RK4 scheme is adequate to allow for a sufficiently large $\Delta t$. Based on numerical experiments, the Courant-Friedrichs-Lewy (CFL) condition \cite{courant1928} is found primarily constrained by the fast parallel electron motion in the case of TCV-X21 (see Section \ref{sec:tcvx21}). Noticing that the term associated with the parallel electron motion is identified in Eq. (\ref{eq:parallelflux}) and that its amplitude scales as $\sqrt{p}$, the largest normalized timestep, $\Delta \hat{t}$, limited by the parallel electron motion can be estimated by

\begin{align} \label{eq:cfl}
  \Delta \hat{t} \lesssim 2.82 \sqrt{\frac{\hat{m}_e}{\hat{\tau}_e}} \frac{\pi \min (\hat R )}{N_\varphi \sqrt{N_{v_\parallel}}},
\end{align}
\\
where we assume that $N_{\mu} \ll N_{v_\parallel}$ small (which is typically the case for turbulence simulations). As indicated by Eq. (\ref{eq:cfl}), the CFL condition becomes more restrictive as the total number of spectral modes in the parallel direction ($N_{v_\parallel}$) and the number of poloidal planes ($N_\varphi$) increase. It is important to note that Eq. (\ref{eq:cfl}) provides only an estimate of the exact CFL condition. In fact, the stability region of the spectral approach depends on the velocity-space resolution used in the simulation since the maximum amplitude of the terms in the spectral GK Vlasov equation, Eq. (\ref{eq:spectralvlasov}), scales with $N_{v_\parallel}$ and $N_\mu$. For instance, in the case where $N_\mu \gg N_{v_\parallel}$, the terms associated with the mirror force (proportional to $j$) dominate over the parallel electron motion and, thus, set the maximum achievable timestep. A more detailed study of these effects is beyond the scope of the present work.

Finally, we remark that the CFL condition imposed by the velocity-space diffusion due to the LBD collision operator scales linearly with the number of spectral nodes. This can be observed in the first line of Eq. (\ref{eq:lbdvspec}) where the contribution from the collisional diffusion is linearly proportional to $p + 2j$. This contrasts with the grid approach where the perpendicular collisional diffusion often imposes the most severe constraint on the timestep as it scales quadratically with the velocity-space resolution. In this case, the time integration of collisions must be treated more carefully in the grid formulation \cite{ulbl2023}. The weaker (linear) dependence of the collisional diffusion on the velocity-space resolution in the spectral approach is advantageous compared to the grid approach since its constraint on the associated CFL condition is less severe in the former than in the latter case. This results from the fact that the LBD collision operator as a sparse representation in the Hermite-Laguerre polynomial basis. 

\subsection{Boundary conditions}
\label{subsec:boundaryconditions}

Although no velocity-space boundary conditions are necessary in the spectral approach, the spectral GK Vlasov equation, given in Eq. (\ref{eq:vlasov}), must be closed at specific values of $p$ and $j$. Therefore, boundary conditions must be provided for the spectral coefficients for which $p = N_{v_\parallel} - 1$ and $j = N_{\mu} - 1$ (see Figure \ref{fig:grid}). For simplicity, we set these boundary spectral coefficients to zero, which is equivalent to truncating the spectral expansion in Eq. (\ref{eq:faexpansion}) at $p = N_{v_\parallel} - 1$ and $j = N_{\mu} - 1$.

Boundary conditions in configuration space are required for the electromagnetic fields and all spectral coefficients. Similar to the grid approach, the electromagnetic fields are subjected to homogeneous Dirichlet boundary conditions everywhere. We also adopt the same boundary conditions as in the grid approach to set the spectral coefficients on the boundary of the computational domain. More precisely, in the grid approach, the distribution function at the boundaries is assumed to be a Maxwellian distribution function with constant temperature and density, fixed by the initial profiles (see Section \ref{subsec:simulationsetup}). The same boundary conditions are applied in the spectral approach by using the analytical projection of a local Maxwellian distribution, which is derived in \ref{appendix:properties} and given in Eq. (\ref{eq:npjmaxw}). These boundary conditions result in the flux variables being set to zero, while the state variables remain non-zero. It should be noted that these boundary conditions are not based on physical principles but have been developed to ensure numerical robustness. Further developments in the spectral approach could include the implementation of more advanced boundary conditions, such as sheath boundary conditions \cite{loizu2012}. 

\section{Verification using the Method of Manufactured Solutions}
\label{sec:mms}

In this section, we verify the numerical implementation of the spectral Vlasov-Maxwell system, given in Eqs. (\ref{eq:spectralvlasov}) and (\ref{eq:spectralmaxwells}), in \verb|GENE-X| using the Method of Manufactured Solutions (MMS) \cite{roache2002}. The MMS approach has been widely used in fluid turbulence codes \cite{riva2014,stegmeir2019} and applied to the grid implementation of \verb|GENE-X| \cite{michels2021}. The MMS verification aims to test the correct numerical implementation of the spectral approach as a whole by conducting a convergence study against known solutions, which are referred to as manufactured solutions. In the absence of an analytical solution to the spectral GK Vlasov-Maxwell system, a modified system is solved instead for the MMS verification. More precisely, the modified GK spectral Vlasov-Maxwell system that we consider can be written as follows

\begin{subequations} \label{eq:gksystemmodified}
\begin{align}
L_V \circ  \hat{\mathcal{N}}_\alpha^{pj}   = S_V, \\
L_Q \circ \hat \phi_1 = S_{Q}, \\
L_A \circ \hat A_{1 \parallel } = S_{A}, \\
L_O\circ  \partial_t \hat A_{1 \parallel } = S_{O}.
\end{align}
\end{subequations}
\\
In Eq. (\ref{eq:gksystemmodified}), the operators $L_V$, $L_Q$, $L_A$, and $L_O$ represent the integro-differential operators associated with the collisionless part of the GK Vlasov equation ($L_V$), the quasineutrality condition ($L_Q$), Ampere’s equation ($L_A$), and Ohm’s law ($L_O$). These operators explicitly depend and act on the spectral coefficients and electromagnetic fields. On the right-hand side of Eq. (\ref{eq:gksystemmodified}), the source terms $S_V$, $S_Q$, $S_A$, and $S_O$ are added such that an analytical solution of the modified GK spectral Vlasov-Maxwell system can be obtained. These analytical solutions, known as the manufactured solutions, are denoted by $\hat{\mathcal{N}}_\alpha^{pj (M)}$, $\hat{\phi}^{(M)}$, $\hat{A}_{1 \parallel }^{(M)}$, and $\partial_t \hat{A}_{1 \parallel }^{(M)}$. These manufactured solutions are arbitrary analytical functions. Once prescribed, the sources $S_V$, $S_Q$, $S_A$, and $S_O$ can be explicitly determined such that Eqs. (\ref{eq:gksystemmodified}) are fulfilled for the manufactured solutions. 

In principle, any arbitrary function can be chosen for the manufactured solutions to compute the sources in Eq. (\ref{eq:gksystemmodified}). However, the manufactured solution must be selected to verify the implementation of the spectral GK Vlasov-Maxwell system effectively. Therefore, we choose the manufactured solutions for the spectral coefficients, $\hat{\mathcal{N}}_\alpha^{pj (M)}$, and for $\hat{\phi}_1^{(M)}$ to be given by

\begin{subequations} \label{eq:mmssol}
\begin{align} \label{eq:npjmms}
\hat{\mathcal{N}}_\alpha^{pj (M)} & = \cos(p + 2j) \nonumber \\
& \times \left(a \sin^2\left( \pi (\hat r - \hat r_{\text{min}}) / (\hat r_{\text{max}}- \hat r_{\text{min}})\right) \right. \nonumber \\
& \left. \times \sin^2 \left( \theta \right) \cos^2 \left( \varphi \right) \cos^2\left( \omega \hat t \right)+ b \right),
\end{align}
\begin{align} \label{eq:fieldmms}
\hat \phi^{(M)}_1  & =  \sin\left( 2 (\hat r - \hat r_{\text{min}}) / (\hat r_{\text{max}}- \hat r_{\text{min}})\right) \nonumber \\
& \times \sin \left( 2 \theta \right) \cos^2 \left( \varphi \right) \cos^2\left( \omega \hat t \right),
\end{align}
\end{subequations}
\\
respectively. We choose $a = 0.95$ and $b = 0.05$. In addition, we set $\hat{A}_{\parallel 1}^{(M)} = \hat \phi^{(M)}_1 $ and $\partial_t \hat{A}_{\parallel 1}^{(M)} = \partial_t \hat \phi^{(M)}_1 $. In Eqs. (\ref{eq:npjmms}) and (\ref{eq:fieldmms}), $\hat{r}$ is the normalized flux surface label (with maximal and minimal values $\hat{r}_{\text{max}}$ and $\hat{r}_{\text{min}}$, respectively), $\theta$ and $\varphi$ are the geometrical poloidal and toroidal angles, and $\omega$ is the real frequency of the manufactured solutions. We remark that the time and spatial dependence of the manufactured solutions given in Eq. (\ref{eq:mmssol}) represents the typical structure of a ballooning mode, while the prefactor $\cos(p + 2j)$ reflects the oscillatory behavior of the amplitude of the spectral coefficients. Given the manufactured solutions in Eq. (\ref{eq:mmssol}), the sources appearing in Eq. (\ref{eq:gksystemmodified}) can be explicitly evaluated using a computer algebra software.

Following Ref. \cite{michels2021}, we perform the MMS verification in three geometries: slab, circular, and toroidal. The details of the magnetic field and safety factor for each geometry can be found in \cite{michels2021}. For the MMS verification, we consider a spectral resolution of $(N_{v_\parallel}, N_\mu) = (4,2)$, which reflects the typical minimal resolution for physically meaningful turbulence simulations. We impose the manufactured solution given in Eq. (\ref{eq:npjmms}) to the spectral coefficients $\hat{\mathcal{N}}_\alpha^{pj}$. The simulation domain extends from $\hat{r}_{\text{min}} = 0$ to $\hat{r}_{\text{max}} = 1$ for the slab and from $\hat{r}_{\text{min}} = 0.5$ to $\hat{r}_{\text{max}} = 0.8$ for the circular and toroidal geometries. In all cases, we consider $\omega = 2\pi$. To perform the MMS verification, we solve the modified system given in Eq. (\ref{eq:gksystemmodified}) in time using RK4, with the manufactured solutions given in Eqs. (\ref{eq:npjmms}) and (\ref{eq:fieldmms}) as initial conditions. At $\hat{t} = 0.625$, we measure the $L_p$-errors, defined by $\norm{fg- g^{(M)}}_p / \norm{g}_p$, where $\norm{g}_p$ is defined by $\norm{g}_2  = \int dV g^2$ for $p=2$ and $\norm{g}_\infty = \max \abs{g}$ for $p = \infty$. Here, $g$ and $g^{(M)}$ are a function (e.g., $\hat{\phi}_1$) and its manufactured solution (e.g., $\hat{\phi}_1^{(M)}$). We note that the time $\hat{t} = 0.625$ is chosen to avoid divergence of the error when the manufactured solution given in Eq. (\ref{eq:mmssol}) approaches zero. The simulations are repeated using five different resolutions. More precisely, the resolution in the poloidal plane and the number of poloidal planes are increased by a factor of two between each case, i.e., we decrease the grid-spacing, $\hat \Delta RZ = 0.0368 / 2^{i}$, and the spacing between successive poloidal planes, $\Delta \varphi = 2\pi / (12 \times 2^{i})$, for $i = 0, 1, 2, 3, 4$. The timestep $\Delta \hat{t}$ is adjusted such that $\Delta \hat{t} = 0.0125 / 2^{i}$. The MMS verification is applied to the GK spectral Vlasov-Maxwell system without collisions and hyperdiffusion (see Section \ref{subsec:hyperdiffusionandissipation}). The exact manufactured solutions given in Eqs. (\ref{eq:npjmms}) and (\ref{eq:fieldmms}) are used as Dirichlet boundary conditions.

\begin{figure}[h]
  \centering
  \includegraphics[scale = 0.5]{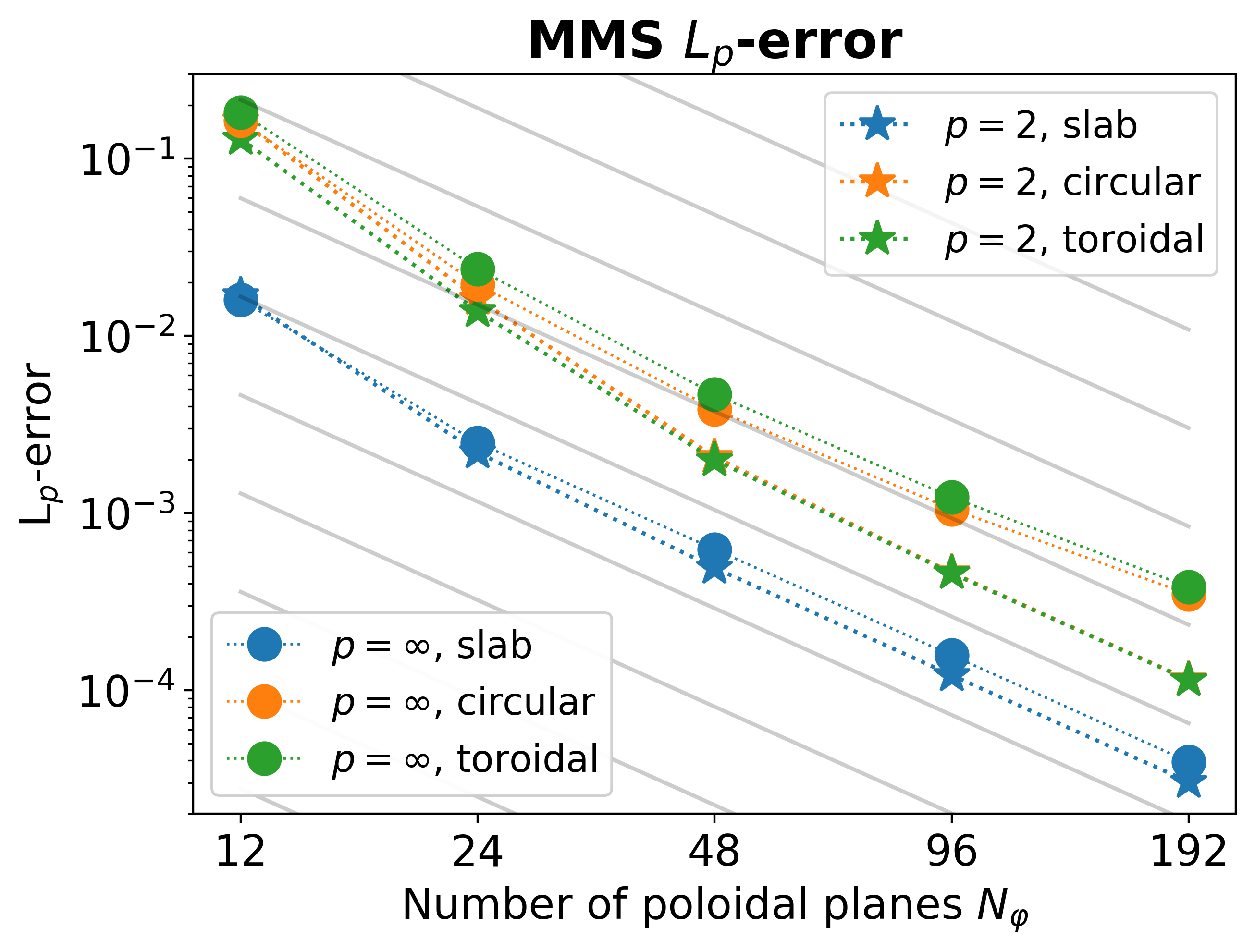}
  \caption{MMS $L_p$-errors of the ion density $\hat n_i = \hat{\mathcal{N}}^{00}_i$  for $p=2$ (star markers) and $p=\infty$ (circular markers) as a function of the number of poloidal planes $N_\varphi$ (as a proxy for the $4$D increase of resolution) in the slab (blue), circular (orange), and toroidal (green) geometry. The second-order of reference is shown by the gray lines.}
  \label{fig:mmserror}
  \end{figure}

  \begin{figure}[h]
  \centering
  \includegraphics[scale = 0.4]{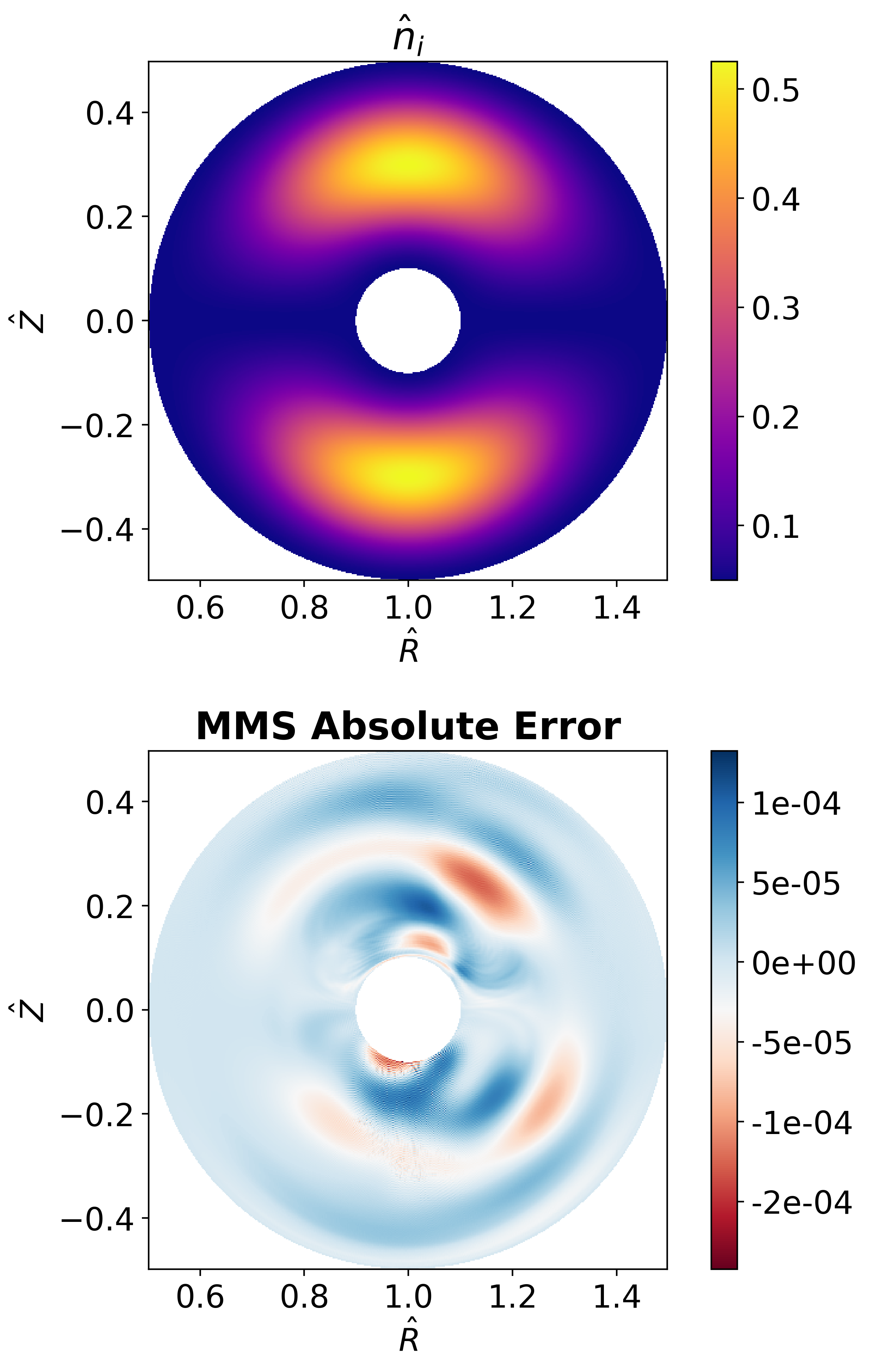}
  \caption{Normalized ion density $\hat n_i$ (top) and MMS absolute error $\hat n_i  - \hat n_i^{(M)}$(bottom) plotted in the $\varphi = \pi$ poloidal plane and at $\hat t = 0.625$ in the toroidal geometry with $N_\varphi = 192$.}
  \label{fig:mmssnapshot}
  \end{figure}

The results of the MMS verification are shown in Fig. \ref{fig:mmserror} where the MMS $L_p$-errors of the normalized ion density, $\hat{n}_i$ are plotted as a function of the number of poloidal planes ($N_\varphi$) for the different geometries. As observed, the MMS errors decrease with second-order accuracy in all geometries at sufficiently high resolution (for $N_\varphi \gtrsim 48$), which is consistent with the dominant error arising from the second-order accuracy of the Arakawa scheme used in \verb|GENE-X| \cite{michels2021}. The faster convergence observed for $N_\varphi \lesssim 48$ is attributed to the largest contribution to the numerical error from the parallel derivative, which is computed using the fourth-order finite difference in the FCI scheme. It should be noted that comparable convergence is observed for the other spectral coefficients and for the electromagnetic fields, but not presented here.

Finally, a snapshot of the normalized ion density and the associated MMS absolute error is displayed in Fig. \ref{fig:mmssnapshot}, obtained at $\hat{t} = 0.625$ in toroidal geometry with $N_\varphi = 192$. We observe that the absolute error has a small amplitude. However, grid scale oscillations appear in the absolute error due to the absence of hyperdiffusion. The convergence observed in Fig. \ref{fig:mmserror} and the small absolute error displayed in Fig. \ref{fig:mmssnapshot} provide confidence in the correct numerical implementation of the spectral GK Vlasov-Maxwell system in \verb|GENE-X|.

\section{First spectral turbulence simulations of TCV-X21}
\label{sec:tcvx21}

This section presents the first spectrally accelerated GK turbulence simulations using \verb|GENE-X|, focusing on the TCV-X21 reference case \cite{oliveira2022}, an experimental scenario in the TCV tokamak \cite{coda2019}. The scenario features an L-mode Ohmically heated Deuterium plasma in a lower-single-null configuration, designed for edge and SOL turbulence code validation. More precisely, the toroidal magnetic field was reduced to lower computational costs, and the electron density was kept low to minimize the impact of neutrals. The TCV-X21 reference case \cite{oliveira2022} has been widely used for the validation of turbulence fluid simulation codes, such as \verb|GBS| \cite{giacomin2021}, \verb|GRILLIX| \cite{stegmeir2019}, and \verb|TOKAM3X| \cite{tamain2016}. This validation exercise has been recently extended to GK simulations using the \verb|GENE-X| code \cite{ulbl2023}.

Unlike the previous fluid code validations \cite{oliveira2022}, the GK simulations reported in Ref. \cite{ulbl2023} show that the interaction between trapped and passing electrons, mediated by collisions, is crucial for accurately predicting experimental measurements, in particular the electron temperature profile. This kinetic mechanism is sensitive to velocity-space features such as the trapped-and-passing boundary, which Braginskii-like fluid codes cannot capture. Therefore, it is essential to assess how well the spectral approach considered in this work can reproduce these findings from grid simulations of the TCV-X21 reference case, where trapped electron dynamics play a dominant role. In this section, we compare the OMP profiles of density, electron and ion temperatures, and radial electric field, with those from grid simulations of \verb|GENE-X| reported in Ref. \cite{ulbl2023}. All the data taken from the grid simulations of \verb|GENE-X| can be found Ref. \cite{tcvx21zenodo} and Ref. \cite{ulbl2023phd}. While only the OMP profiles are considered in this work, a more detailed analysis of turbulence characteristics predicted by the spectral simulations will be presented in a future publication.

This section is structured as follows. The simulation setup of the TCV-X21 reference case is outlined in Section \ref{subsec:simulationsetup}, snapshots from the spectral simulations are discussed in Section \ref{subsec:simulationresults}, and the OMP profiles are compared with grid simulations in Section \ref{subsec:ompprofiles}. 

\subsection{Simulation setup}
\label{subsec:simulationsetup}

 Due to the design of the numerical implementation of the spectral approach in \verb|GENE-X| (see Section \ref{sec:numericalimplementation}), only the velocity-space resolution needs to be adjusted with respect to Ref. \cite{ulbl2023}. We, thus, only recall the most important points and refer to Ref. \cite{ulbl2023} for more details. We use $N_\varphi = 32$ poloidal planes and an uniform grid spacing of $\Delta RZ \simeq 1.23$~mm, resulting in a total of $N_{RZ} = 200657$ grid points per poloidal plane. The normalization parameters (see \ref{appendix:normalization}) are $\Tref = 20$ eV, $\nref = 10^{19 }$ m$^{-3}$, $\Bref = 0.929$~T, and $\Lref = 0.906$~m, implying $\csref \simeq 43770$~km/s, $\betaref = 9.33 \times 10^{-5}$, and $\rhoref \simeq 0.5$~mm. Finally, we consider a Deuterium plasma with $\hat m_{\mathrm{i}} = 2$ and $\hat m_{\mathrm{e}} \simeq 1/1830$. 

To assess the performance of the spectral approach, we consider different spectral resolutions by increasing the number of spectral coefficients, $N_{v_\parallel}$ and $N_\mu$, in the parallel and perpendicular directions, respectively. More precisely, we use $(N_{v_\parallel}, N_\mu) = (4,2)$, $(6,2)$, $(8,4)$, and $(16,8)$. To satisfy the CFL condition given in Eq. (\ref{eq:cfl}), we use a normalized timestep of $\Delta \hat{t} = 4 \times 10^{-4}$ for the lowest spectral resolution and $\Delta \hat{t} = 3.5 \times 10^{-4}$ for the highest spectral resolution of $(16,8)$. A constant damping coefficient of $K_\parallel = 0.1$ is applied (see Eq. (\ref{eq:landauclosure})). 

%Similar to the grid simulations, a floor density and temperature of 0.25 of the reference values, $\nref$ and $\Tref$, are used in the LBD collision operator since the spectral scheme is not positivity-preserving.

\begin{figure}
\centering
\includegraphics[scale = 0.5]{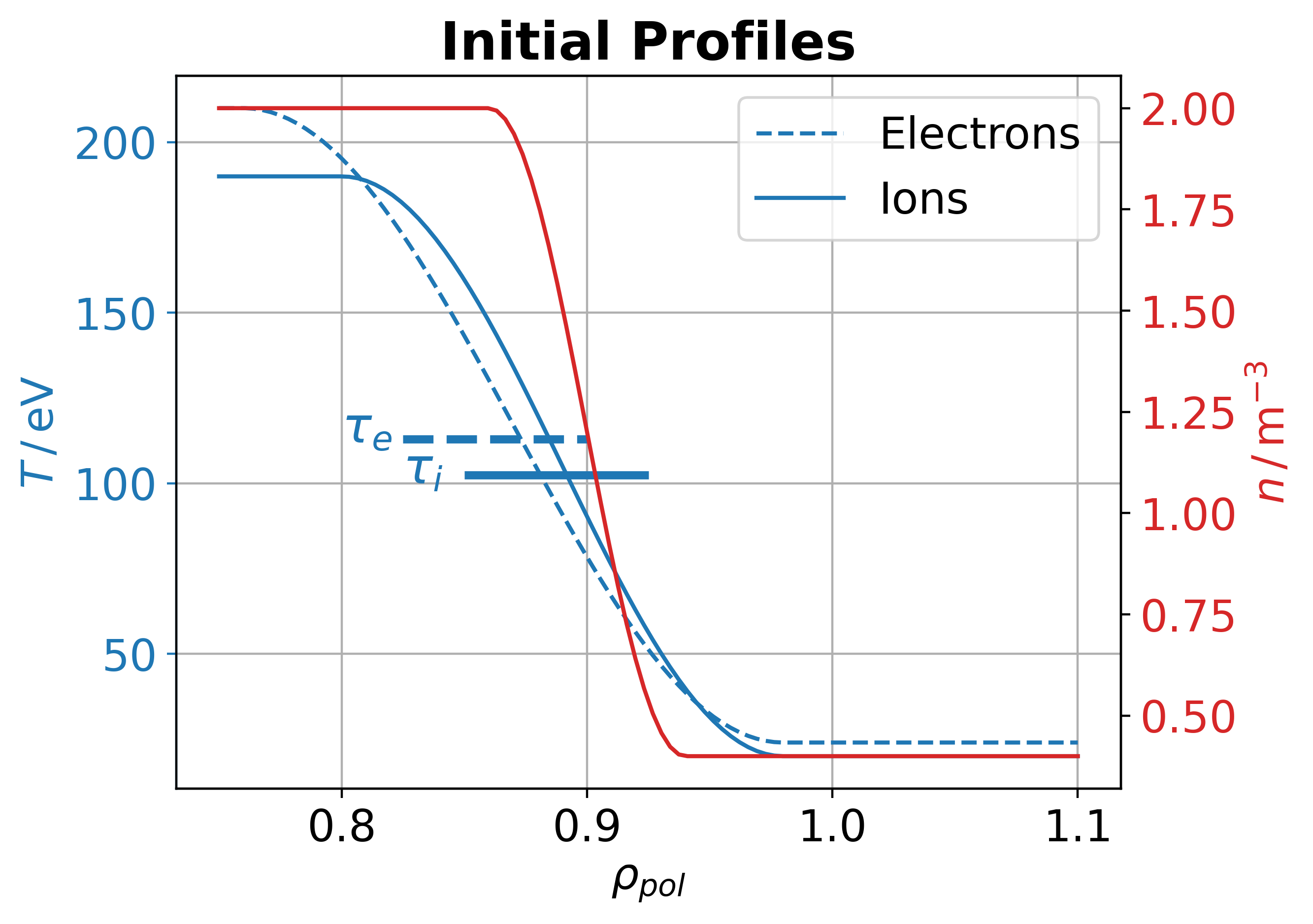}
\caption{Initial density $n$ (red) and temperature $T_{e/i}$ (blue) profiles for the electrons (dashed) and ions (solid) as a function of the poloidal flux surface label $\rho_{\mathrm{pol}}$. The density profiles are the same for both species. The horizontal blue lines represent the values of the scaled temperature, $\tau_\alpha$, such that the condition, Eq. (\ref{eq:crittaua}), is fulfilled.}
\label{fig:initialprofiles}
\end{figure}

 Similarly to the grid simulations, the electromagnetic fields, $\phi_1$ and $A_{1 \parallel}$, are set to zero as initial conditions throughout the computational domain. On the other hand, the spectral coefficients are initialized using the analytical spectral projection of a local Maxwellian distribution, as detailed in Eq. (\ref{eq:npjmaxw}), which is evaluated using the initial density $n_\a$ and temperature $T_\a$ initial profiles depicted in Fig. \ref{fig:initialprofiles}. We remark that these initial profiles are identical to those used in Ref. \cite{ulbl2023}. Finally, based on the $T_\alpha$ initial profiles, the scaled temperatures $\tau_\alpha$ are adjusted such that the criterion given in Eq. \ref{eq:crittaua} is satisfied (see \ref{appendix:properties}). This yields $\tau_e = 114$~eV and $\tau_i = 102.5$~eV. It should be noted that these values are not unique and are typically chosen based on trial and error to ensure stable initial simulations.

The spectral simulations presented in this work were performed on different CPU-based supercomputers, including the Cobra ($2 \times 20$ cores per node with Intel Xeon Gold at $2.4$ GHz) and Raven ($2 \times 36$ cores per node with Intel Xeon IceLake at 2.4 GHz) supercomputers at the Max-Planck Computing and Data Facility (MPCDF), and the A3 partition ($2 \times 24$ cores per node with Intel Xeon 8160 at 2.10 GHz) of the Marconi supercomputer at Cineca. The typical requirement for each simulation is approximately $32$ nodes, with a runtime of a few days, for approximately $0.1$ MCPUh in total, depending on the machine and resolution. This computational cost is much less than the one associated with grid simulations, which typically is of the order of a few MCPUh. In Section \ref{sec:computational}, we provide a detailed account of the computational resources required for these spectral simulations and discuss the associated speed-up compared to the grid approach.

\subsection{Saturation and snapshots}
\label{subsec:simulationresults}

\begin{figure}[h]
\centering
\includegraphics[scale = 0.5]{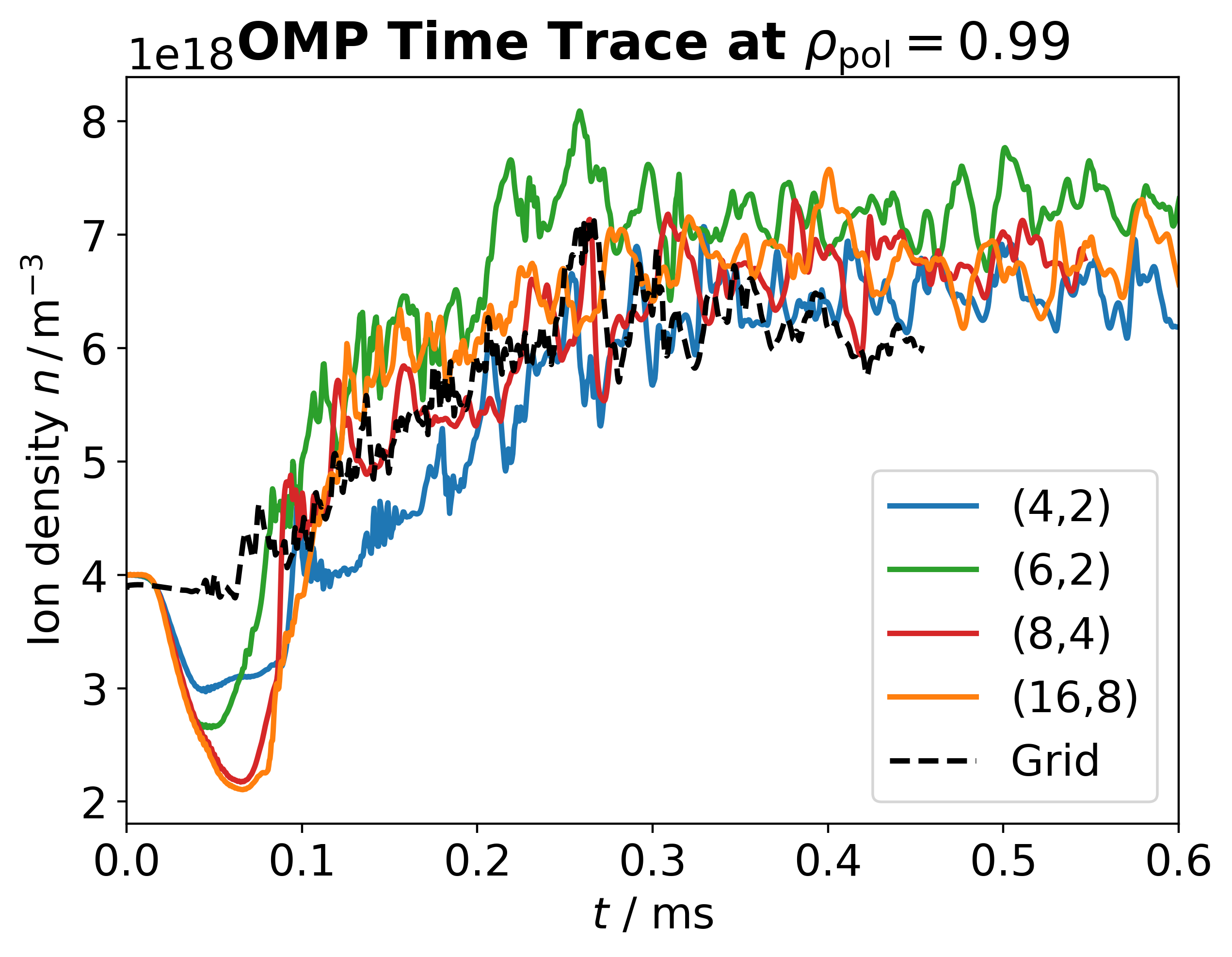}
\caption{Time traces of the toroidally averaged density close to the separatrix at the OMP ($\rho_{\mathrm{pol}} = 0.99$) obtained from the spectral (colored line) and (black dashed line) grid simulations. Data taken from \cite{tcvx21zenodo}. In all cases, a quasi-steady state is reached for $t \gtrsim 0.45$~ms. }
\label{fig:omptimetrace}
\end{figure}

Starting from the initial conditions detailed in Section \ref{subsec:simulationsetup}, the spectral simulations are run until a quasi-steady state is achieved. To monitor the quasi-steady state, we compute the toroidal average of the density and temperatures near the separatrix at the OMP position, around $\rho_\mathrm{pol} \simeq 0.98$. Fig. \ref{fig:omptimetrace} shows the resulting time traces from the different spectral simulations. As observed, a quasi-steady state is achieved after a simulation time of $t \gtrsim 0.4$~ms in all cases. We remark that the simulation time required in the spectral simulations to achieve a quasi-steady state is of the order of the one obtained with the grid simulations performed using the LBD collision operator model \cite{ulbl2023}, as shown by the black dashed line in Fig. \ref{fig:omptimetrace}. Similar time traces are observed for the electron and ion temperatures but are not shown here. We note that the different times associated with the turbulence onset (at $t \lesssim 0.1$~ms) may be attributed to the possible sensitivity of the linear growth rates due to the different spectral resolutions used in the simulations \cite{frei2023}.

\begin{figure}[h]
\centering
\includegraphics[scale = 0.35]{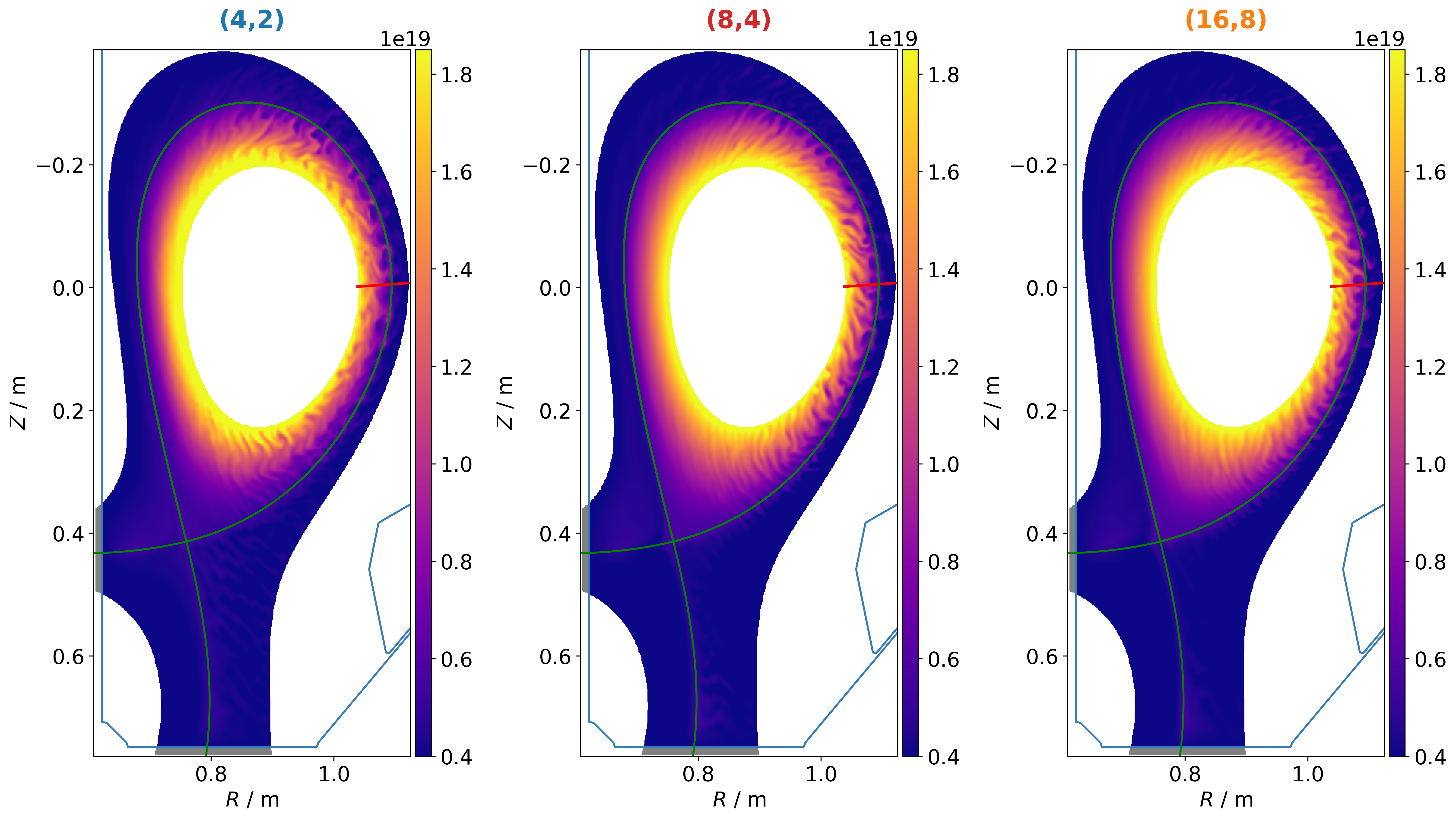}
\caption{Snapshots of the ion density $ n_{\mathrm{i}}$ in the poloidal plane $ \varphi = \pi $ at $ t = 0.5$~ms, obtained with increasing spectral resolutions (from left to right). The green and red solid lines indicate the position of the separatrix and the OMP measurement line, respectively.}
\label{fig:densitysnapshots}
\end{figure}

Fig. \ref{fig:densitysnapshots} displays snapshots of the ion density obtained with increasing resolution. The snapshots are evaluated at the $\varphi = \pi$ poloidal plane at $t = 0.5$~ms. From the inspection of Fig. \ref{fig:densitysnapshots}, it is evident that turbulence structures are present in the region of closed field lines ($\rho_\mathrm{pol} < 1$) and extend into the region of open field lines ($\rho_\mathrm{pol}  > 1$). Furthermore, the turbulence activity is stronger in the low-field side than in the high-field side, indicating the ballooning nature of the underlying instabilities. While no discernible differences are observed between the simulations with $(N_{v_\parallel}, N_{\mu}) > (4,2)$, a poloidal asymmetry can be noted near the inner boundary when $(N_{v_\parallel}, N_{\mu}) = (4,2)$. This poloidal asymmetry is associated with the presence of artificially undamped parallel flows arising from the lack of a damping mechanism related to the parallel dynamics such as, e.g., Landau damping, which is not well resolved when $N_{v_\parallel} \lesssim 4$. As a result, these parallel flows lead to large-scale geodesic acoustic mode (GAM)-like oscillations. To remove these spurious oscillations, an increase of the parallel resolution (e.g., $N_{v_\parallel} = 6$) or an increase of the damping coefficient $K_\parallel$ (see Eq. (\ref{eq:landauclosure})) can limit their appearance and, thus, suppresses the poloidal asymmetry observed near the inner boundary. This can be observed in Fig. \ref{fig:densitysnapshots} when increasing the parallel resolution to $N_{v_\parallel} = 8$.

\begin{figure}
\centering
\includegraphics[scale = 0.4]{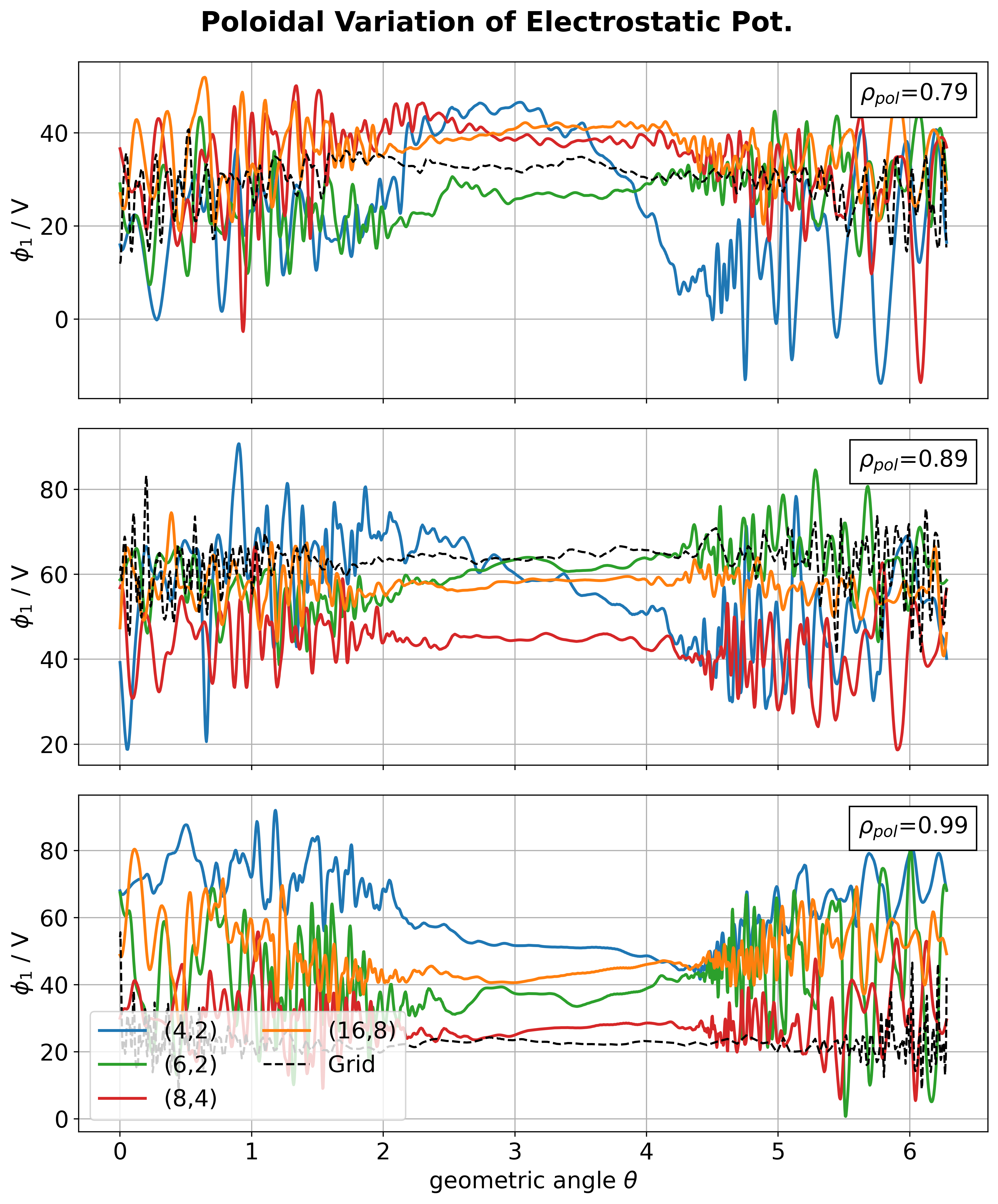}
\caption{Poloidal variation of the electrostatic potential, $\phi_1$, as a function of the geometrical angle $\theta$ on different flux surfaces obtained using different spectral resolutions (colored lines). The dashed black lines also show the poloidal variation of $\phi_1$ from the grid simulations for comparison.}
\label{fig:poloidalvarphi}
\end{figure}

Fig. \ref{fig:poloidalvarphi} illustrates the instantaneous poloidal variation of the normalized electrostatic potential, $\hat{\phi}_1$, on different flux surfaces evaluated in the $\varphi = \pi$ poloidal plane. The electrostatic potential is plotted as a function of the geometrical angle $\theta$ for the spectral and grid simulations. The low-field side is located near $\theta = \pi$, while the high-field side corresponds to $\theta = 0$ and $2\pi$. Similar to the grid simulations, the electrostatic potential exhibits more pronounced variations on the low-field side, with much smaller fluctuations on the high-field side, clearly indicating the ballooning nature of the turbulence. Overall, we observe good qualitative behavior of the instantaneous electrostatic potential structures despite the presence of spurious poloidal asymmetry near the inner boundary around $\rho_{\mathrm{pol}}  = 0.79$ in the $(4,2)$ simulation. We finally remark that the grid simulations reported in Ref. \cite{ulbl2023} utilize a linearized polarization in the quasi-neutrality condition given in Eq. (\ref{eq:poisson}), which corresponds to the Boussinesq approximation in fluid models \cite{ross2018}. While the effects of this approximation on turbulence properties in geometries with X-points are beyond the scope of this study, we do not expect this assumption to significantly alter the turbulent structures \cite{ross2018} displayed in Fig. \ref{fig:densitysnapshots}. 

\begin{figure}[h]
\centering
\includegraphics[scale = 0.4]{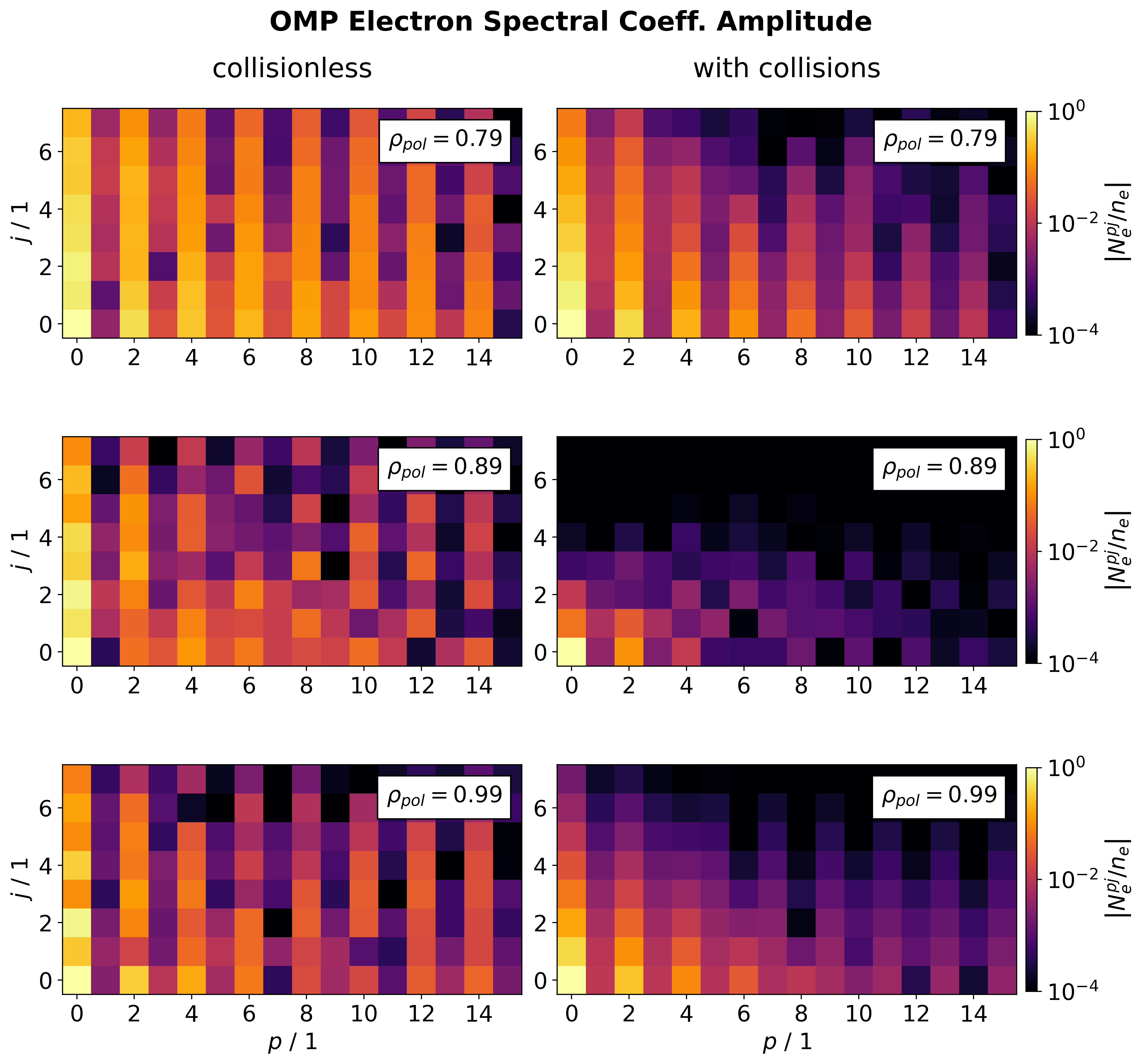}
\caption{Normalized amplitude of the electron spectral coefficients without (left column) and with (right column) collisions. The spectral coefficients are evaluated at quasi-steady state in the $\varphi = \pi$ poloidal plane and at different $\rho_{\mathrm{pol}}$ at the OMP (see Fig. \ref{fig:densitysnapshots}). Here, $(N_{v_\parallel}, N_\mu)= (16,8)$.}
\label{fig:finalnpj}
\end{figure}

Finally, Fig. \ref{fig:finalnpj} displays the normalized amplitude of the electron spectral coefficients in the case of $(N_{v_\parallel }, N_\mu) = (16,8)$, evaluated on different flux surfaces at the OMP in the poloidal plane $\varphi = \pi$ at quasi-steady-state. In addition to the results from the simulations with collisions, the electron spectral coefficients from a collisionless simulations are shown for comparison.

Focusing on the case with collisions, we observe that the amplitudes of the electron spectral coefficients decrease by at least four orders of magnitude in the parallel direction (along $p$). In contrast, a slower decay is observed in the perpendicular direction (along $j$). The broader electron spectrum near the separatrix ($\rho_{\mathrm{pol}}  \simeq 0.99$) and near the inner boundary ($\rho_{\mathrm{pol}}  \simeq 0.79$), in comparison to that evaluated at $\rho_{\mathrm{pol}}  \simeq 0.89$, is a consequence of the value of the electron scaled temperature, being $\tau_\mathrm{e} = 114$~eV, which is nearly equal to the local electron temperature around $\rho_{\mathrm{pol}}  \simeq 0.89$ (see Fig. \ref{fig:ompte}). Indeed, the spectral expansion performed in Eq. (\ref{eq:faexpansion}) has better converge properties if the typical width of the Hermite-Laguerre basis is set to the local fluid temperature $T_\alpha$, i.e., when $\tau_\alpha$ is close to $T_\alpha$ (see \ref{appendix:properties}). On the other hand, when $\tau_\alpha$ deviates from $T_\alpha$, as in the far SOL and near the inner boundary, the spectral convergence is less optimal, leading to the broader spectrum.

By comparing the spectral amplitude without and with collisions (shown in the left and right columns of Fig. \ref{fig:finalnpj}), we observe that collisional effects damp the higher-order spectral coefficients, leading to a faster decay in both the parallel and perpendicular directions. Consequently, simulations with collisions require a lower spectral resolution to achieve convergence compared to the collisionless case \cite{frei2023}, highlighting the advantage of this approach in describing plasma dynamics when collisions dominate. Overall, Fig. \ref{fig:finalnpj} indicates that a resolution of $(N_{v_\parallel}, N_\mu) = (16,8)$ is sufficient for well-resolved spectral simulations, as confirmed below for both the collisional and collisionless cases.

\subsection{Outboard midplane profiles}
\label{subsec:ompprofiles}

\begin{figure}[h]
\centering
\includegraphics[scale = 0.5]{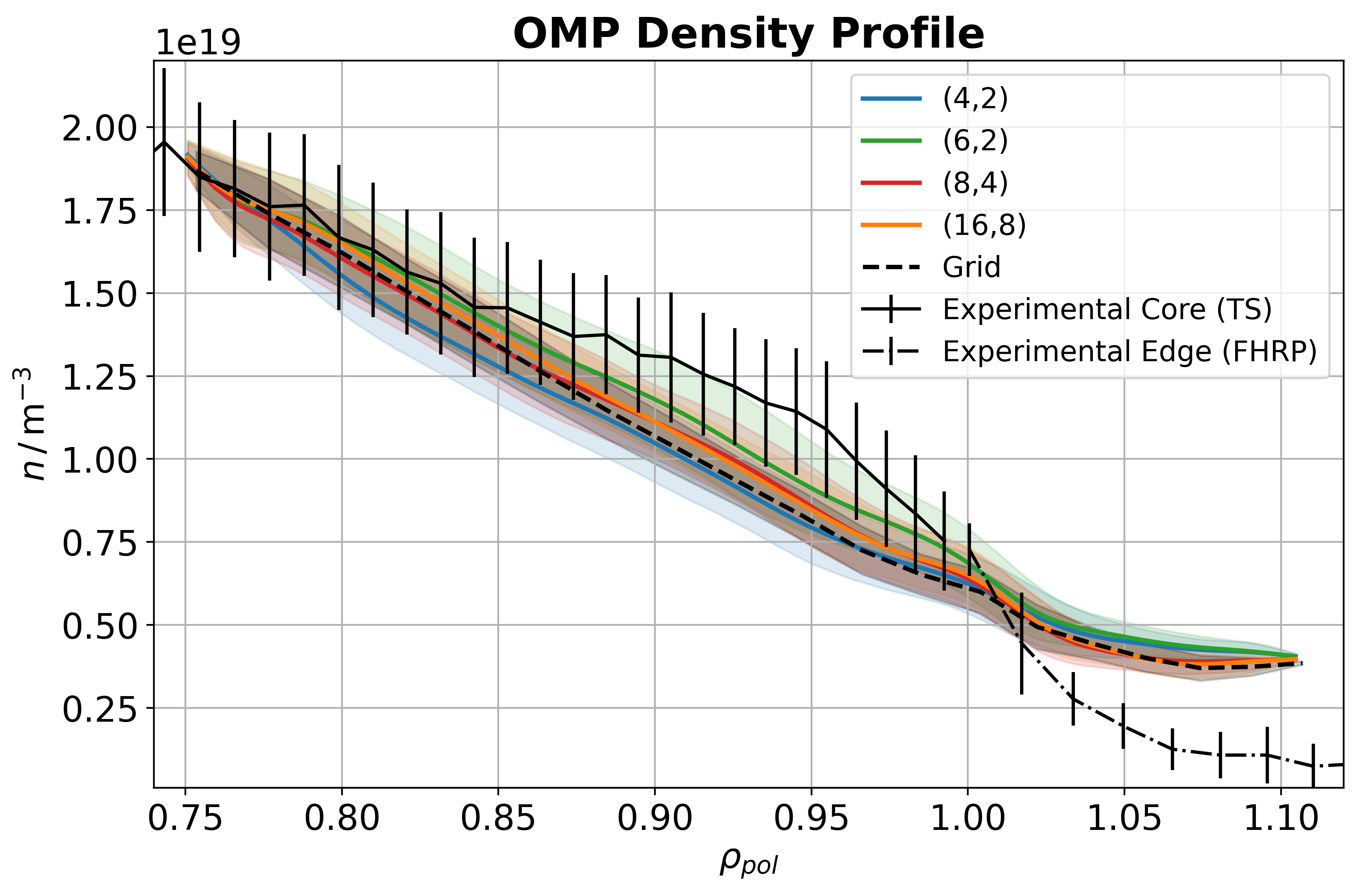}
\caption{OMP ion density $n_{\mathrm{i}}$ profiles obtained using the spectral simulations (colored lines) with different spectral resolution $(N_{v_\parallel}, N_\mu)$. The OMP profiles are computed by performing a toroidal and time average on the data interpolated on the OMP line of measurements (see red line in Fig. \ref{fig:densitysnapshots}) over $0.1$~ms. For comparison, the dashed black line shows the OMP profiles obtained from the grid simulation (data taken from Ref. \cite{tcvx21zenodo}). The shaded areas are the standard deviations of the OMP profiles. The solid and dotted black lines show the experimental measurements from TS and FHRP, respectively.}
\label{fig:ompdens}
\end{figure}

We now evaluate the OMP profiles of density, temperatures, and radial electric field from the spectral simulations and compare them with those obtained from the grid simulations reported in Ref. \cite{ulbl2023}. The data of these simulations can be found in Ref. \cite{tcvx21zenodo}. We also include experimental measurements \cite{oliveira2022} for the density and electron temperature from Thomson scattering (TS) diagnostics in the core region and from the fast reciprocating probe (FHRP) in the SOL region. These data can be found in Ref. \cite{tcvx21zenodoOliveira}.

To obtain the OMP profiles, a toroidal and time average is performed on the data interpolated at the OMP position, indicated by the red solid line in Fig. \ref{fig:ompdens}. The time average is calculated over a $0.1$~ms period in all simulations at quasi-steady state. It is worth noticing that the OMP profiles from the grid simulations are obtained similarly. In Fig. \ref{fig:ompdens}, the OMP profiles of the density obtained from the spectral simulations are shown and compared with the OMP profile obtained from the grid simulations. The results from the grid and spectral simulations are all in excellent agreement (as well as the standard deviations). However, the lowest spectral resolution slightly underestimates the density profile in the edge region where $\rho_{\mathrm{pol}} \lesssim 0.9$. Finally, the deviations from the experimental measurements can be attributed to the absence of neutrals \cite{zholobenko2021b} in the \verb|GENE-X| code.

\begin{figure}[h]
\centering
\includegraphics[scale = 0.5]{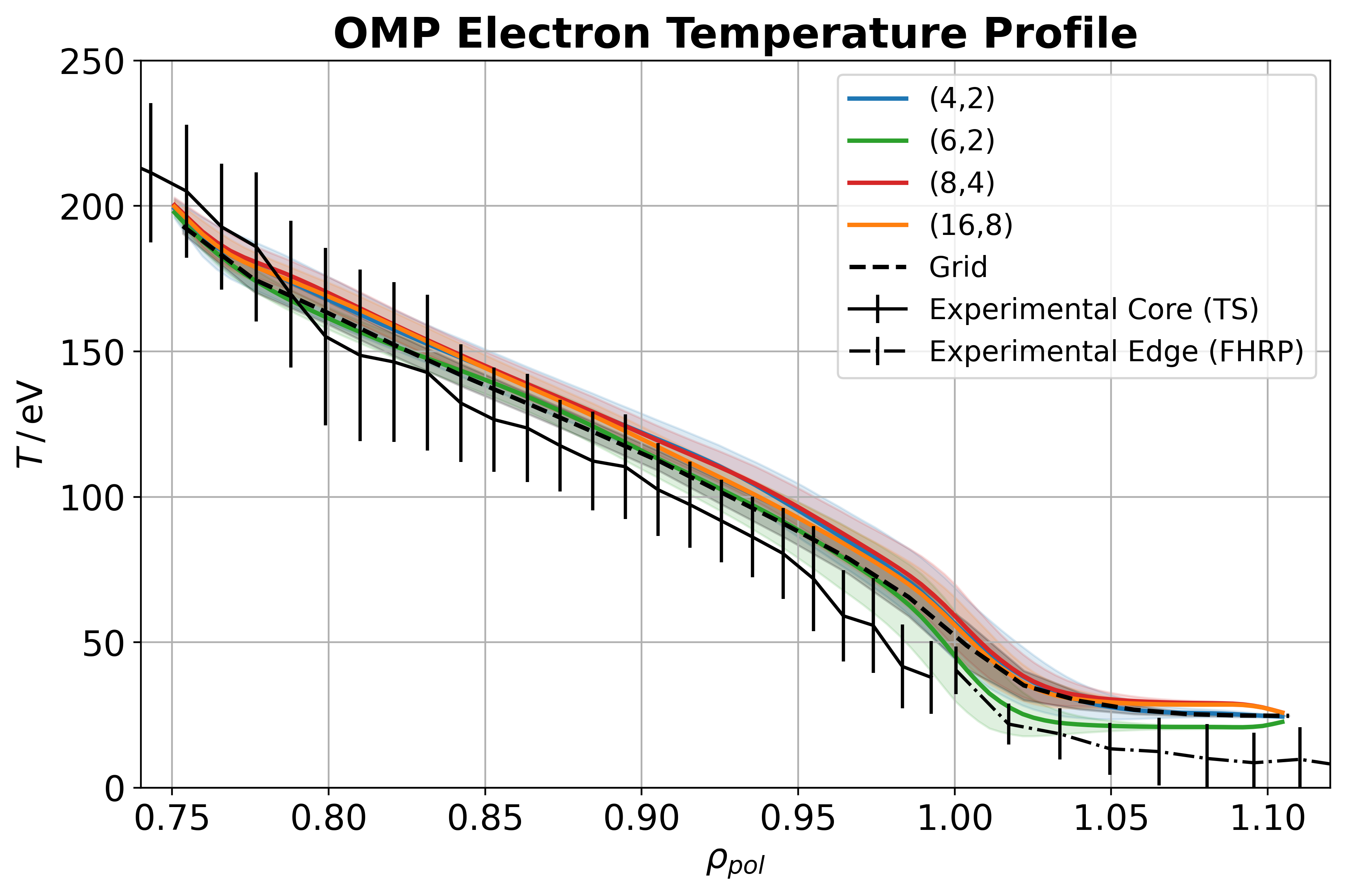}
\caption{Same as Fig. \ref{fig:ompdens} for the OMP electron temperature $T_{\mathrm{e}}$ profiles.}
\label{fig:ompte}
\end{figure}

We now focus on the OMP profile of the electron temperature $ T_{\mathrm{e}} $. In the TCV-X21 validation of \verb|GENE-X| \cite{ulbl2023}, it was found that TEMs are the main drivers of turbulence, explaining the discrepancies between fluid model predictions and experimental measurements of electron temperature profiles in the edge region. 
More precisely, it has been established that the collisional exchange between perpendicular energy, carried by trapped electrons and ignored in fluid simulations, and parallel energy, carried by passing electrons, is a crucial mechanism for accurately matching the experimental $T_{\mathrm{e}}$ profile in the TCV-X21 reference case. Therefore, similarly to the density OMP profile, we evaluate the OMP $T_{\mathrm{e}}$ profiles in the different spectral simulations and compare them with those from the simulation and experimental measurements. The results are displayed in Fig. \ref{fig:ompte} and show that the OMP $T_{\mathrm{e}}$ profiles from the spectral simulations closely match the one predicted by the grid simulation, even with the lowest resolution $(N_{v_\parallel}, N_\mu) = (4,2)$. 
This is an unexpected result given that trapped electron physics is generally sensitive to velocity-space resolution, which can be challenging when using a spectral approach due to fine scale features near the trapped and passing boundary \cite{frei2023}. Nevertheless, the good agreement observed between the OMP profiles obtained from the grid and spectral simulations in Fig. \ref{fig:ompte}, in addition to the weak dependence on spectral resolution, suggests that the spectral approach implemented in this work provides an accurate description of the cooling of the trapped electrons by collisional effects, key physical ingredient to predict the electron temperature profile.

\begin{figure}[h]
\centering
\includegraphics[scale = 0.5]{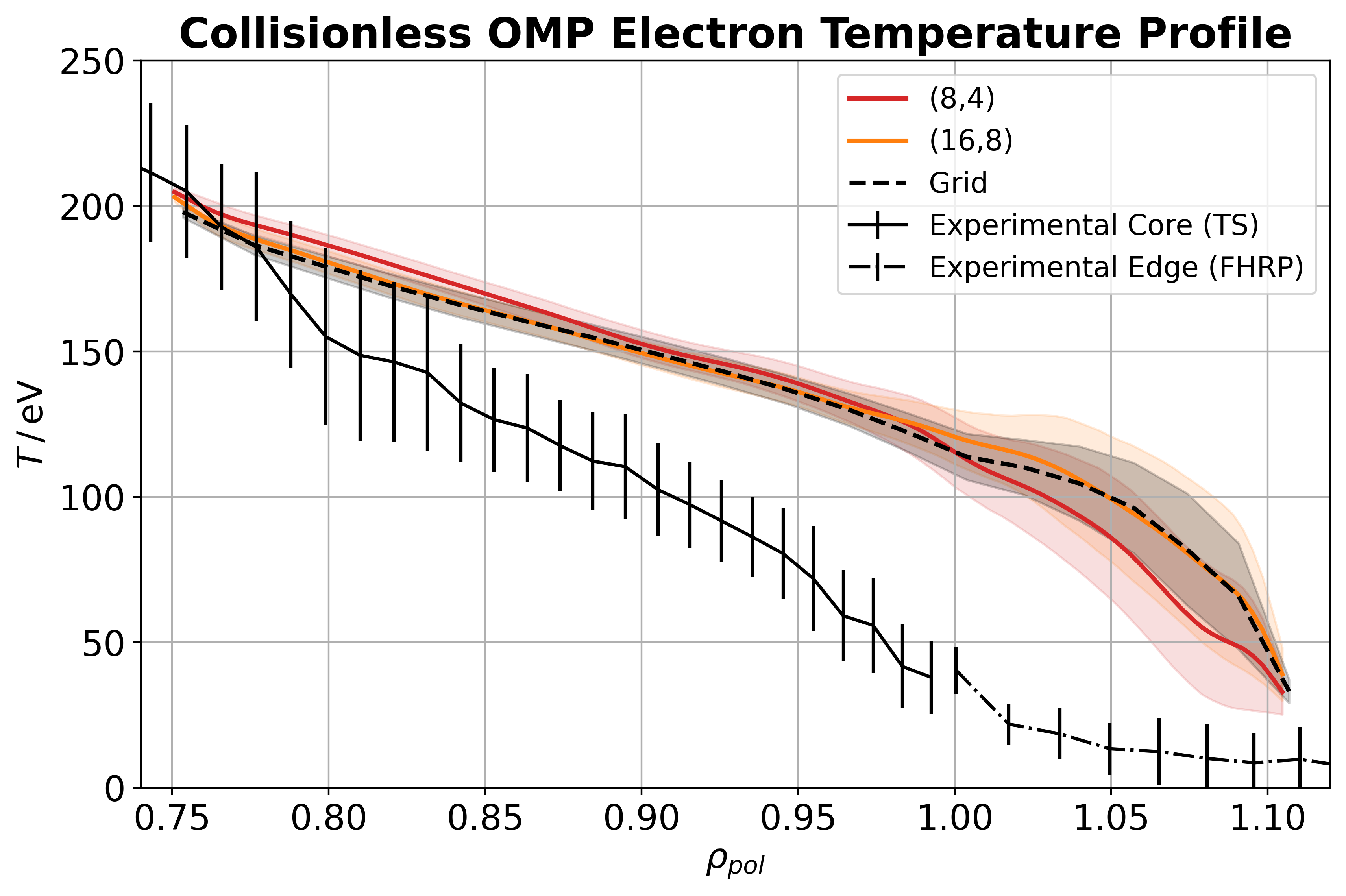}
\caption{Same as Fig. \ref{fig:ompte}, but in the collisionless limit. In this case, only the spectral resolutions $(8,4)$ and $(16,8)$ are considered. The experimental measurements are shown to ease the comparison with Fig. \ref{fig:ompte}.}
\label{fig:omptenocoll}
\end{figure}

To further evaluate the ability of the spectral approach to capture trapped electron physics, we perform two spectral simulations in the collisionless limit (by neglecting the LBD collision operator on the right-hand side of Eq. (\ref{eq:spectralvlasov})) with $(8,4)$ and $(16,8)$ spectral coefficients. In this case and consistent with the collisionless grid simulation \cite{ulbl2023}, a longer simulation time of $t \gtrsim 0.7$~ms, compared to the collisional case, is required to reach a quasi-steady state. The collisionless OMP $T_{\mathrm{e}}$ profiles are plotted in Fig. \ref{fig:omptenocoll} and compared with the grid simulation \cite{ulbl2023}. Similarly to the collisional case, we observe a remarkably good agreement between the spectral and grid simulations even in this collisionless case. However, this agreement is achieved with a higher spectral resolution. Indeed, while no significant difference is observed in the simulations with $(8,4)$ and $(16,8)$ in the presence of collisions, we observe a noticeable difference between these two cases, in particular in the SOL. At the same time, they both yield comparable profiles in the edge region. More precisely, the $(16,8)$ simulations closely match the collisionless grid simulation everywhere.
On the other hand, deviations are observed in the SOL region with a spectral resolution $(8,4)$. This higher sensitivity on the spectral resolution without collisions illustrates the improved convergence of the spectral approach when collisional effects are important. Based on these observations, Figs. \ref{fig:ompte} and \ref{fig:omptenocoll} show that the spectral approach implemented in \verb|GENE-X| can capture well TEM-dominated turbulence in both the collisional and collisionless limits. Furthermore, it can be observed that the level of fluctuations (illustrated by the shaded colored areas) is comparable in both the spectral and grid simulations. However, we postpone a more detailed analysis to a future publication. 

\begin{figure}[h]
\centering
\includegraphics[scale = 0.5]{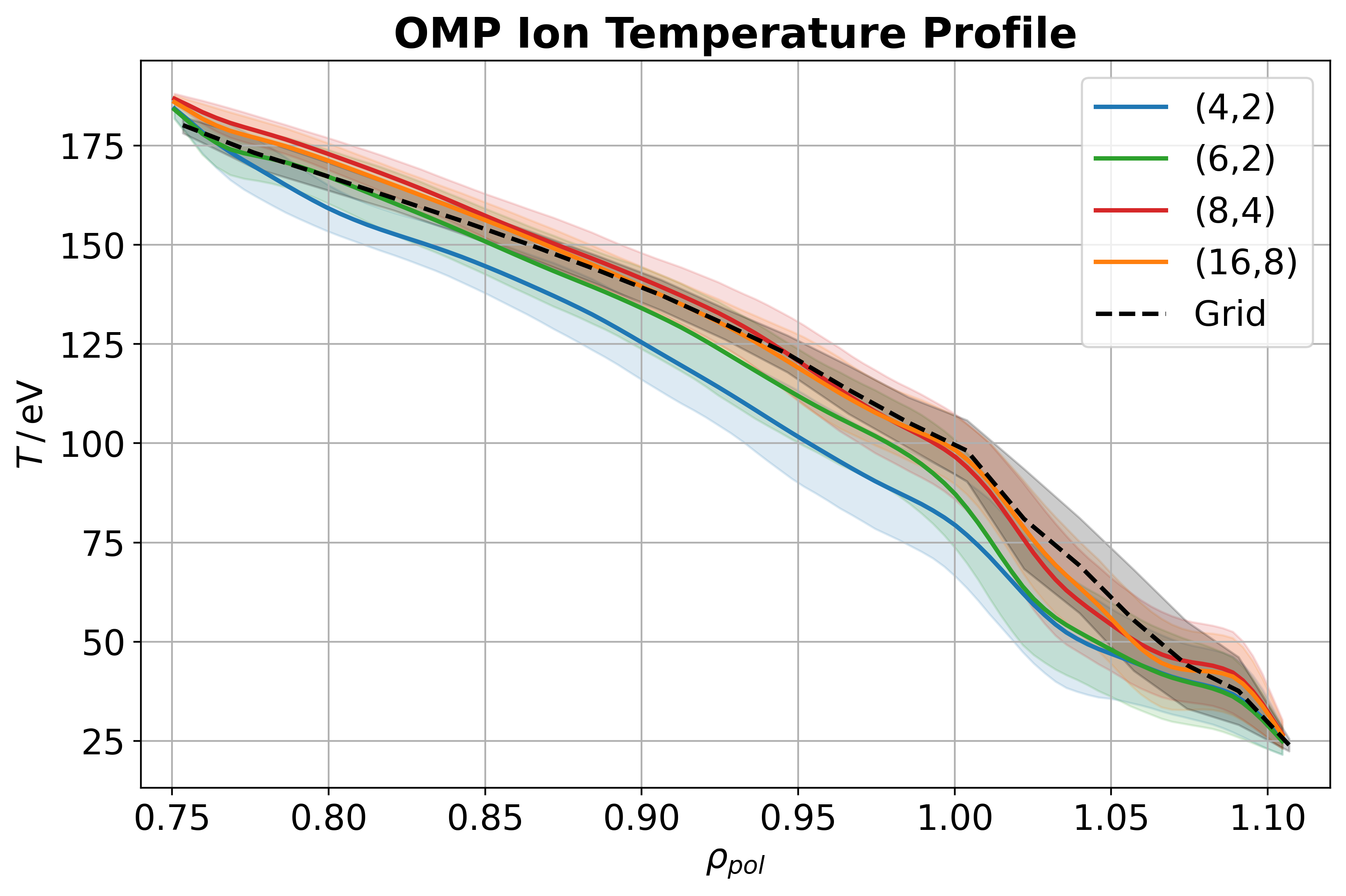}
\caption{Same as Fig. \ref{fig:ompte} for the OMP ion temperature $T_{\mathrm{i}}$ profiles. No experimental measurements are available in this case.}
\label{fig:ompti}
\end{figure}

We now turn to the OMP profiles of the ion temperature $T_{\mathrm{i}}$. The profiles obtained from the spectral simulations are plotted in Fig. \ref{fig:ompti}. There are no experimental measurements available in this case. Unlike the density and electron temperature profiles, the OMP $T_{\mathrm{i}}$ profile changes with spectral resolution. While the lowest resolution $(4,2)$ underestimates $T_{\mathrm{i}}$ near the separatrix, a good agreement with the grid simulation is obtained with $(N_{v_\parallel}, N_\mu) = (8,4)$ and higher. In particular, increasing $N_{v_\parallel}$ (e.g., from $(4,2)$ to $(6,2)$) improves the $T_{\mathrm{i}}$ profile in the edge region, while the profile near the separatrix remains almost unchanged. Further increasing the resolution in the perpendicular direction, i.e. increasing $N_{\mu}$, improves the overall agreement. The sensitivity to the parallel resolution demonstrates the influence of the parallel ion dynamics in the edge region, while the dependence on $N_\mu$ highlights the importance of resolving the perpendicular ion temperature dynamics, $T_{\perp \mathrm{i}}$ (see Eqs. (\ref{eq:mom2da}), near the separatrix and in the SOL. Finally, we notice that the agreement observed in the far SOL ($\rho_{\mathrm{pol}} \gtrsim 1.05$) is due to the Dirichlet boundary condition and should therefore not be considered as physical. 

\begin{figure}[t]
\centering
\includegraphics[scale = 0.5]{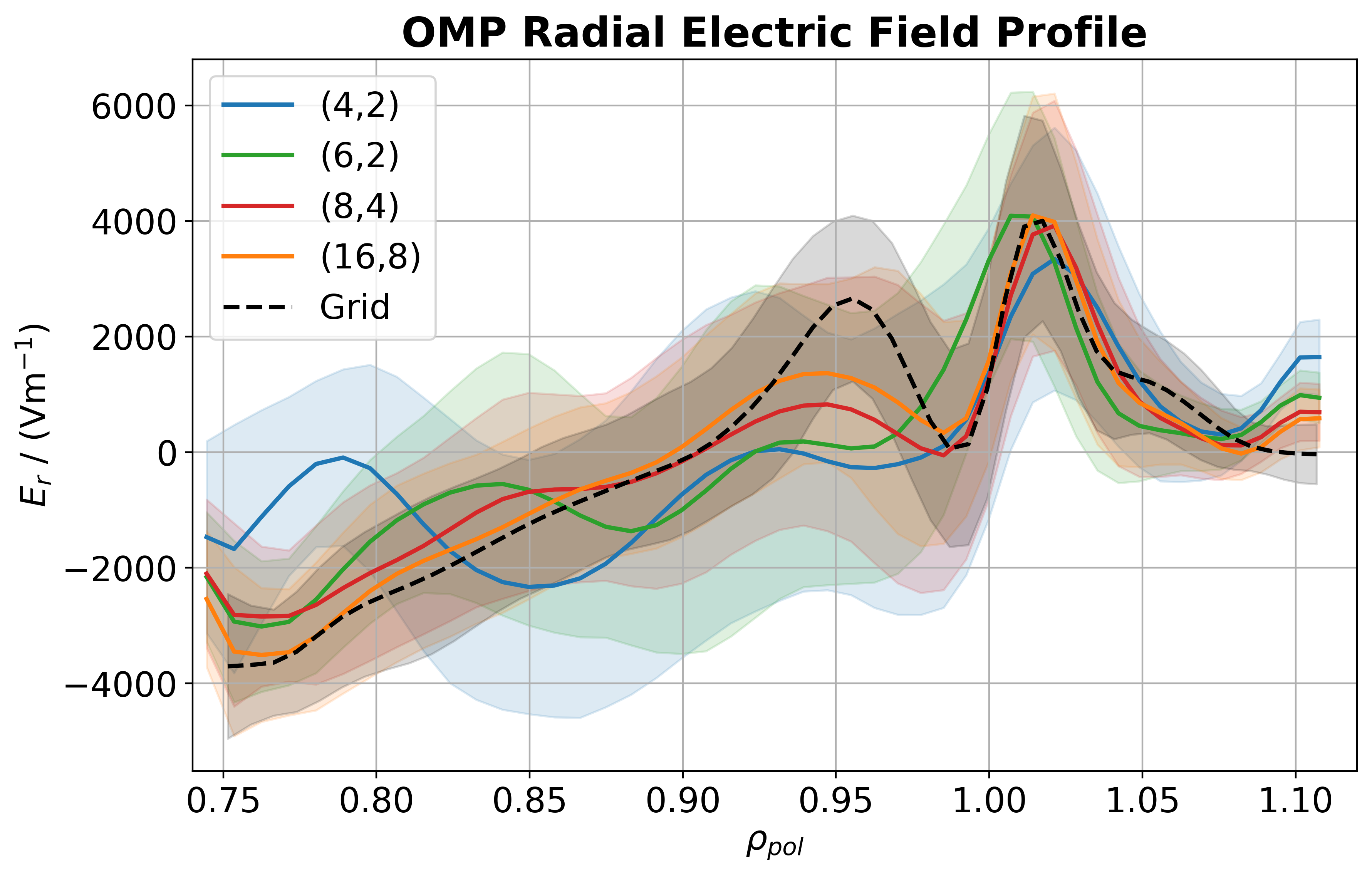}
\caption{Same as Fig. \ref{fig:ompdens} for the OMP radial electric field, $E_r$, profiles. Data taken from Ref. \cite{ulbl2023phd}.}
\label{fig:ompEr}
\end{figure}

Finally, we consider the OMP profile of the radial electric field, given by $ E_r = - \nabla \phi_1 \cdot \nabla r $, which is shown in Fig. \ref{fig:ompEr}. In the SOL, the radial electric field is positive and increases towards the separatrix before decreasing to negative values in the edge region. It is observed that while the highest resolutions give an almost monotonic $E_r$ profile in the edge for $\rho_{\mathrm{pol}}  \lesssim 0.92$, the cases with $(4,2)$ and $(6,2)$ show an oscillating $E_r$ profile in the edge. These local maxima and minima in the radial electric field result from the poloidal asymmetries observed in Fig. \ref{fig:densitysnapshots} and in Fig. \ref{fig:poloidalvarphi}. Increasing $N_\mu$ improves the $E_r$ profile at the edge and suppresses these oscillations. This dependence of the $E_r$ profile on the spectral resolution is not observed in the SOL. Despite these spurious oscillations at low resolutions, a good qualitative agreement with the grid simulation \cite{ulbl2023phd} is obtained. Finally, the radial electric field vanishes towards the inner and outer boundaries due to the homogeneous boundary conditions. Thus, the OMP profile of $E_r$ close to the boundaries should not be considered physical.

\section{Performance analysis and computational cost}
\label{sec:computational}

In this section, we evaluate the performance and computational cost of the spectral approach and evaluate the associated speed-up with respect to the grid implementation of \verb|GENE-X|. We start by performing a roofline analysis in Section \ref{subsec:roofline}. Then, in Section \ref{subsec:computationcost}, we report the computational costs of the TCV-X21 spectral simulations presented in Section \ref{sec:tcvx21} and assess the speed-up enabled by the spectral approach in Section \ref{subsec:speedup}.

\subsection{Roofline analysis}
\label{subsec:roofline}

The roofline analysis \cite{williams2009} helps to identify potential bottlenecks and inefficiencies affecting the overall performance of both the spectral and grid implementations in \verb|GENE-X|. This performance model enables to measure and compare the maximal achievable performance (in GFlops/s) of each operator used for the various computations involved in the collisional GK Vlasov-Maxwell system. More precisely, we analyze the following operators: the static and dynamic components of the right-hand side (rhs) of the GK Vlasov equation (referred to as rhs static and rhs dynamic, respectively), the LBD collision operator (coll lbd), and the computation of fluid moments (mom), which are used for solving the GK Maxwell equations. These operators account for most of the computation time per timestep in \verb|GENE-X|. 

\begin{figure}[h]
\centering
\includegraphics[scale = 0.5]{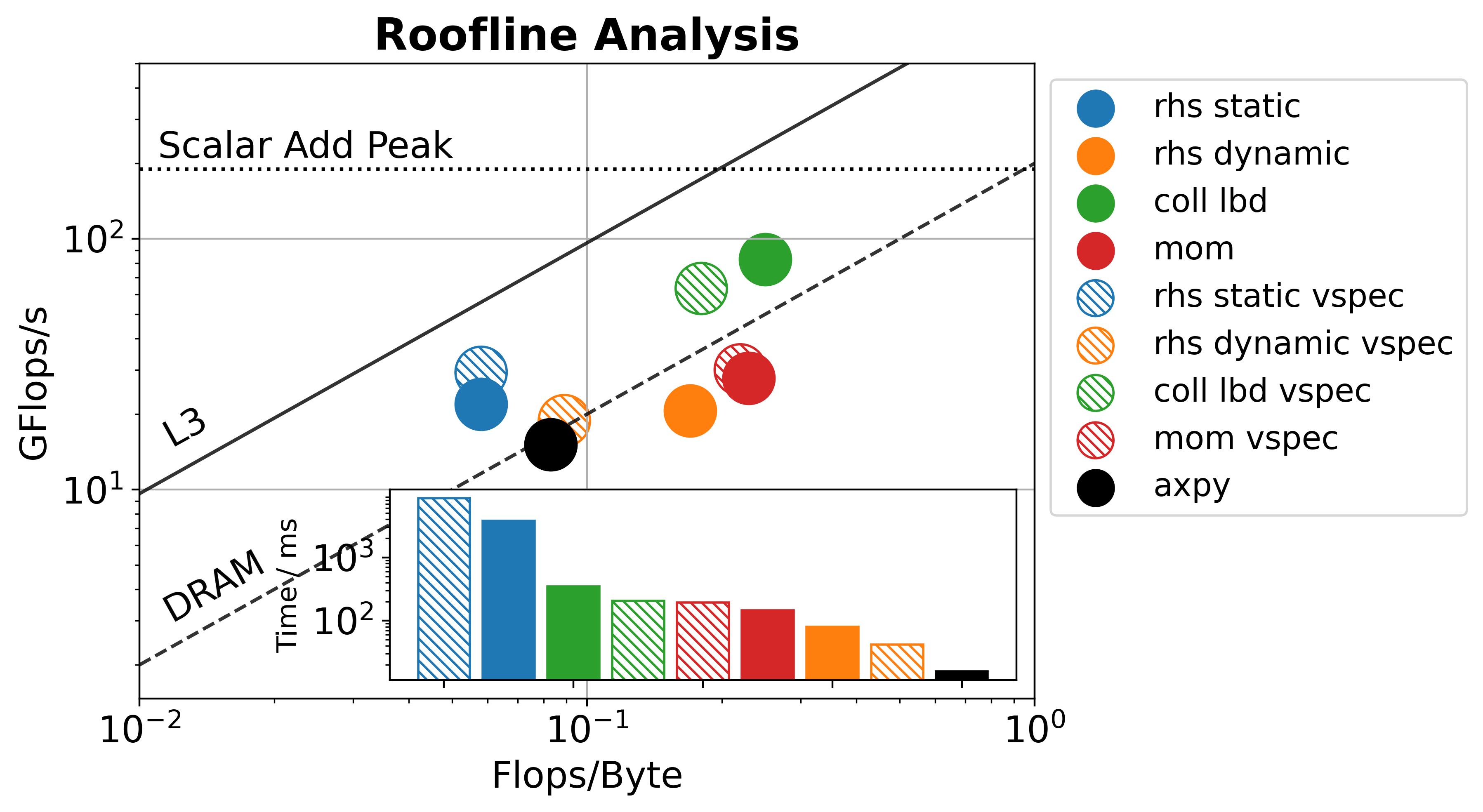}
\caption{Roofline analysis, showing the performance (GFlops/s) as a function of the arithmetic intensity (Flops/Byte), of the spectral (denoted by vspec) and grid operators associated with the computations of the static (blue) and dynamic (orange) part of the Vlasov equation, the LBD collision operator (green), and fluid moments (red). The solid colors show the grid operators and the hatched colors show the spectral operators. The roofline analysis for the axpy operator (black) is also shown for comparison. The inset shows the elapsed time (in ms) of each operator on a logarithmic scale. The DRAM (dashed) and L3 (solid) bandwidth and scalar add peak (dotted) ceilings are shown.}
\label{fig:roofline}
\end{figure}

For the roofline analysis, we select a problem size representative of the typical load on a single compute node in realistic, high-resolution simulations, characterized by $N_{RZ} \simeq 10^6$, $N_{\phi} = 2$, $N_{v_{\parallel}} = 8$, $N_{\mu} = 4$, and $N_{\alpha} = 2$, amounting to approximately $0.13$ billion points. We note that $N_{v_{\parallel}} = 8$ and $N_{\mu} = 4$ typically represents the number of velocity-space grid points distributed per node in a grid-based simulation, while this is comparable to the total number of spectral coefficients used in a spectral simulation. The analysis is performed on the Cobra supercomputer ($2 \times 20$ cores per node with Intel Xeon Gold at $2.4$ GHz). The results, presented in Fig. \ref{fig:roofline}, show the maximum achievable performance of each spectral (vspec) and grid operator as a function of the arithmetic intensity (flops/byte). Additionally, the associated elapsed times are displayed in the inset. As observed, both the grid and spectral implementations exhibit similar maximum performance and arithmetic intensity. However, the dynamic part of the Vlasov equation and the LBD collision operator have lower arithmetic intensity in the spectral implementation because they require fewer arithmetic operations than their grid counterparts. Overall, all operators have low arithmetic intensity, indicating that performance is primarily limited by memory bandwidth. In particular, the dynamic and moment computations are constrained by DRAM bandwidth, while the static and collision computations are limited by L3 cache bandwidth ($962.84$ GByte/s), suggesting better cache reuse in the latter cases. We remark that the static part of the Vlasov equation remains the most time-consuming operator in both the grid and spectral approaches. In conclusion, the roofline analysis reveals that the grid and spectral formulations have comparable computational intensities and elapsed times, and are predominantly memory-bound.

\subsection{Computational cost}
\label{subsec:computationcost}

Table \ref{table:summary} summarizes the computational resources used to carry out the spectral simulations presented in Section \ref{sec:tcvx21}. This table also includes two additional simulations, one spectral and one grid (data taken from \cite{ulbl2023}), performed using a coarser resolution in the poloidal plane, i.e. $\Delta RZ = 3.7 \rhoref$, and with an optimized $\mu$ grid \cite{ulbl2023phd}. These two simulations are used below to evaluate the speed-up achieved by the spectral approach.

Since the simulations presented in Section \ref{sec:tcvx21} were not all run until the same end time, we assess their computational cost based on the time required to reach a simulation time of $ t = 0.5 $~ms, which corresponds to a quasi-steady state (see Fig. \ref{fig:omptimetrace}). The computational cost is then calculated using the average time per timestep listed in Table \ref{table:summary}. The results (in MCPUh) are shown in Fig. \ref{fig:computationalcost}, where the simulations from Section \ref{sec:tcvx21} are represented by solid colored bars. Fig. \ref{fig:computationalcost} indicates that a spectral simulation of TCV-X21 requires between $ 0.05 $ and $ 0.1 $~MCPUh, depending on the spectral resolution. The wall-clock runtime typically ranges from $24$~hours for, e.g., the $(8,4)$ resolution to approximately $50$~hours for the highest $(16,8)$ resolution on the Raven supercomputer. It is worth noting that spectral simulations achieve an average timestep of approximately $2$~s and use between $32$ and $64$ nodes. In comparison, the grid simulations have a much longer time per timestep of $10$ seconds on average with about $256$ nodes. Overall, the four simulations discussed in Section \ref{sec:tcvx21} were completed within a few days on different supercomputers, consuming less than $0.5$~MCPUh in total. 

\subsection{Speed-up of the spectral approach}
\label{subsec:speedup}

\begin{figure}[h]
\centering
\includegraphics[scale = 0.5]{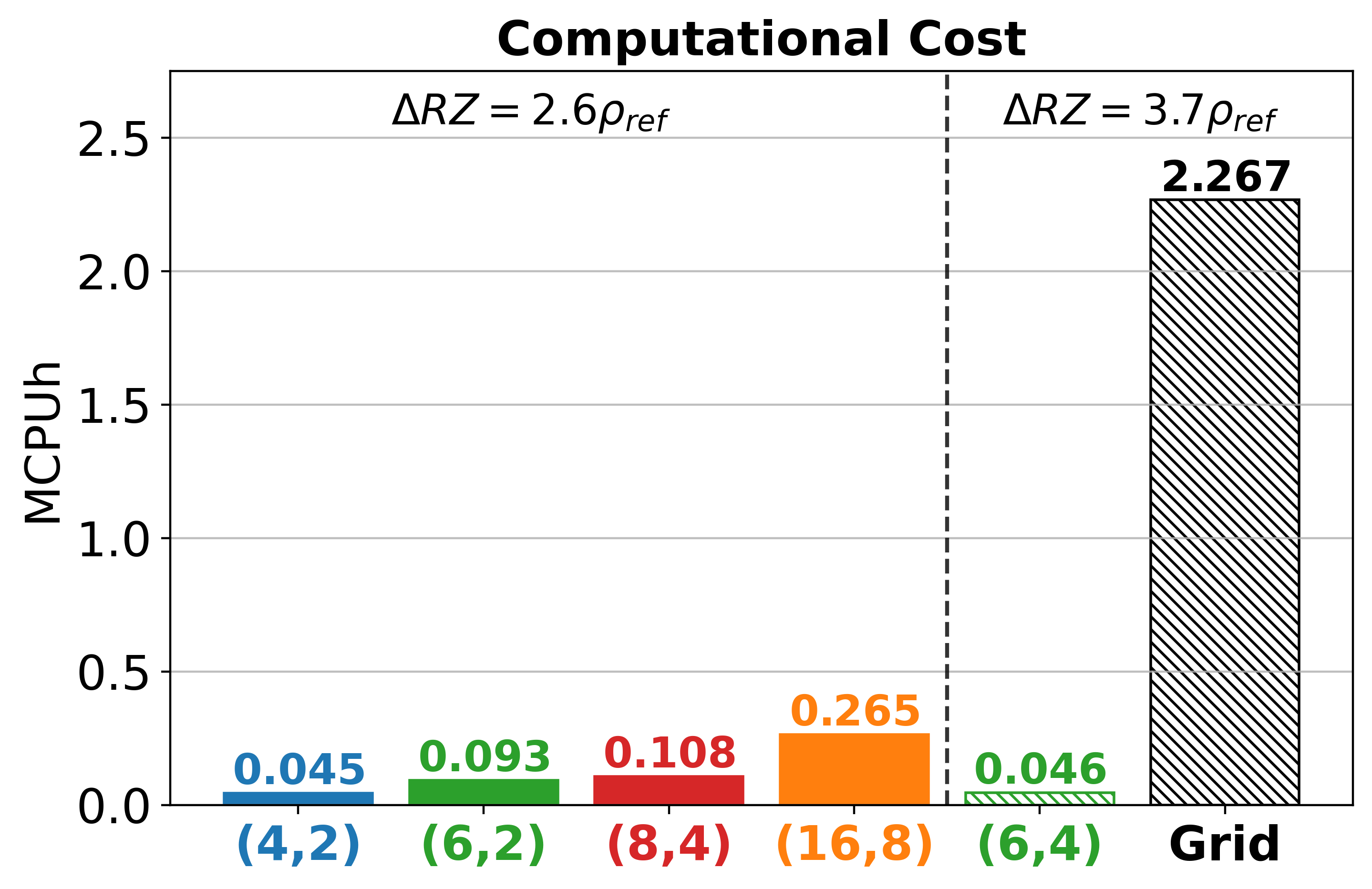}
\caption{Comparison of computational costs (MCPUh) of the spectral simulations performed in Section \ref{sec:tcvx21} with $\Delta RZ = 2.6 \rhoref$ (filled colored bars). For the computational cost comparison, a velocity-space resolution of $(N_{v_\parallel}, N_{\mu}) = (6, 4)$ for the spectral and $(N_{v_\parallel}, N_{\mu}) = (80, 24)$ for the grid approach (taken from Ref. \cite{ulbl2023phd})) are used with a grid-spacing of $\Delta RZ = 3.7 \rhoref $ (dashed colored bars). All the computational costs are calculated based on a simulation time of $t = 0.5$ ms with an average time per timestep taken from Table \ref{table:summary}.}
\label{fig:computationalcost}
\end{figure}

To assess the computational speed-up enabled by the spectral formulation, we compare the coarser grid (data taken from Ref. \cite{ulbl2023phd}) and spectral simulations with $\Delta RZ = 3.7 \rhoref$. This larger grid-spacing, compared to Section \ref{sec:tcvx21}, saves computational resources for this comparison. In addition, we use these simulations rather than those reported in Ref. \cite{ulbl2023} since the former uses an equidistant grid in $v_\perp$, optimizing the number of grid points near $ \mu = 0 $. Consequently, the number of grid points $ N_\mu $ can be reduced compared to Ref. \cite{ulbl2023}, from $60$ to $24$. Therefore, in total, the grid simulation uses a velocity-space resolution of $(N_{v_\parallel}, N_\mu) = (80,24)$. To compute the speed-up of the spectral approach, we use the A3 partition of the Marconi supercomputer for the grid and spectral simulations. The same grid-spacing, $\Delta RZ = 3.7 \rhoref $, and timestep, $ \Delta \hat{t} = 4 \times 10^{-4} $ are utilized in both simulations and are run until $ t = 0.5 $~ms with the same number of timesteps. For the grid simulation, we use $256$ nodes, achieving an average time per timestep of $11$~s. This translates into about $2.2$~MCPUh, as shown in Fig. \ref{fig:computationalcost} by the hatched color bar. For the spectral simulation, a velocity-space of $ (N_{v_\parallel}, N_\mu) = (6,4) $ is chosen (based on the observations from Section \ref{sec:tcvx21}) with $64$ nodes. As a consequence, a much shorter time per timestep of $0.9$~s is achieved, resulting into a runtime of approximately $16$ hours and consuming a total of only $0.046$~MCPUh, as shown in Fig. \ref{fig:computationalcost}. Therefore, the spectral approach provides a substantial speed-up of around $50$ times compared to the optimized grid simulations reported in Ref. \cite{ulbl2023phd}.

To summarize, the spectral approach introduced in this work enables the acceleration of high-fidelity GK edge and SOL turbulence simulations with \verb|GENE-X|, allowing simulations of medium-sized devices (e.g., TCV and AUG) to be completed within a few days on CPU-based supercomputers. However, it is important to note that the speed-up achieved by the spectral approach may vary depending on the plasma scenario and the computational architecture.

\begin{table}[t]
\centering
\begin{tabular}{cccccccccc||c}
$(N_{v_\parallel}, N_\mu)$ & Type & $\Delta RZ / \rhoref$ & Machine & Time per $\Delta \hat  t$ & $\#$ Nodes & $\#$ CPUs \\
\hline
\hline
$\phantom{00}(4,2)$ & S  &  $2.6$ & C &   $\phantom{0}2.1$ s  & $\phantom{0}$32 & $\phantom{0}$1280 \\
$\phantom{00}(6,2)$ &S & $2.6$  &M &  $\phantom{0}1.8$ s & $\phantom{0}$32 & $\phantom{0}$3072 \\
$\phantom{00}(8,4)$ &S & $2.6$  &R  &  $\phantom{0}1.4$ s & $\phantom{0}$64& $\phantom{0}$4608  \\
$\phantom{0}(16,8)$& S & $2.6$ &R  &  $\phantom{0}3.0$ s & $\phantom{0}$64& $\phantom{0}$4608  \\
$\phantom{00}(6,4)$ &S  & $3.7$ &M  &   $\phantom{0}0.9$ s & $\phantom{0}$64 & $\phantom{0}$3072  \\
$(80,24) $& G &  $3.7 $ &M  &   $11.0$ s  & 256 &  12288  \\
\end{tabular}
\caption{Summary of the computational resources used for the spectral (S) simulations, reported in Section \ref{sec:tcvx21}, and grid (G) simulations, data taken from Ref. \cite{ulbl2023phd} with $\Delta RZ = 3.7 \rhoref$, of TCV-X21. The simulations were performed on the Cobra (C), Raven (R), and the A3 partition of the Marconi (M) supercomputers. For each simulation, the velocity-space resolution $(N_{v_\parallel}, N_\mu)$, perpendicular grid-spacing $\Delta RZ$, supercomputer used, average time per timestep $ \Delta \hat t$, and total number of nodes and CPUs are reported.}
\label{table:summary}
\end{table}

\section{Conclusions}
\label{sec:conclusion}

In this work, we have presented the first GK simulations of edge and SOL turbulence accelerated by a spectral method in velocity-space implemented in the \verb|GENE-X| code. This establishes \verb|GENE-X| as the first full-$f$ code capable of performing high-fidelity GK turbulence simulations using a velocity-space spectral approach. % The first turbulence simulations of the TCV-X21 diverted L-mode reference case \cite{oliveira2022} have been successfully carried out and compared with previously validated grid simulations of \verb|GENE-X| \cite{ulbl2023}. Finding an excellent agreement on the OMP profiles dominated by TEM turbulence, the spectral method approach achieves a significant computational speed-up of approximately $50$ compared to the grid approach in the TCV-X21 reference case.

The spectral formulation considered in this work involves expanding the distribution function onto a velocity-space basis of Hermite-Laguerre polynomials with a scaled temperature that efficiently adapts the basis. The electromagnetic and collisional GK Vlasov-Maxwell system was then derived in terms of the spectral coefficients of the distribution function and implemented in \verb|GENE-X|. The object-oriented design of \verb|GENE-X| facilitated the integration of the spectral formulation, leveraging the existing grid infrastructure. The numerical implementation was rigorously verified using the method of manufactured solutions (MMS) in different geometries.

A series of simulations of the TCV-X21 reference case were conducted with an increasing number of spectral coefficients. Focusing on the outboard midplane profiles, the results showed excellent agreement with previous validated grid simulations \cite{ulbl2023}, which were identified to be dominated by TEM turbulence. We remark that the TCV-X21 reference case provides a robust physical scenario for assessing the efficiency of the spectral method in accurately describing edge and SOL turbulence since the dynamics of trapped electrons is sensitive to velocity-space structures. Overall, we observed that the agreement improved with increasing spectral resolution, with the highest resolution, $(N_{v_\parallel}, N_\mu)\sim(16,8)$, producing accurate results in both the collisional and collisionless limits, even in the presence of trapped electrons. Our comparison has shown that a spectral resolution of $(N_{v_\parallel}, N_\mu)\sim(6,4)$ is sufficient to reproduce the results of grid simulations, which typically require a much higher velocity-space resolution. We hypothesize that a similar spectral resolution is adequate for L-mode scenarios in larger machines such as AUG \cite{michels2022}. However, the necessary velocity-space resolution is likely to vary depending on temperature gradients across the domain, especially in H-mode plasmas, and on the specific instabilities present (e.g., TEM, ITG, or KBM \cite{frei2023}). Although this study primarily focused on comparing OMP profiles with grid simulations, future work will extend this comparison to include analyses of turbulence characteristics. Nonetheless, preliminary results show good agreement across all quantities.

The computational cost of the spectral simulations conducted in this work was evaluated and compared with previous grid simulations. We have demonstrated that the spectral formulation significantly reduces the velocity-space resolution, resulting in a substantial speed-up of approximately $50$ times for the TCV-X21 case. While this speed-up may be case specific, it enables high-fidelity GK simulations of devices comparable in size to TCV and AUG \cite{michels2022} to be completed within a few days on current CPU-based supercomputers. This computational efficiency opens new opportunities for studying edge and SOL turbulence in reactor-relevant devices using \verb|GENE-X|.

Despite the significant speed-up achieved by the spectral approach, further developments are necessary to make the spectral approach of \verb|GENE-X| a fully predictive tool for realistic scenarios in reactor-relevant conditions. Future work includes implementing sheath boundary conditions, coupling with neutral models, incorporating FLR effects, and porting to GPU architectures.

%Despite the large speed-up enabled by the spectral approach, further developments are still necessary to make \verb|GENE-X| a fully predictive tool for realistic scenarios in reactor-relevant conditions. This include implementing sheath boundary conditions, coupling with a neutral models, FLR effects, and GPU porting.

\section*{Acknowledgement}

The authors would like to acknowledge Andreas Stegmeir, Wladimir Zholobenko, and Christoph Pitzal for their valuable discussions on the verification of the numerical implementations and their insightful contributions to the physical interpretation of the results. We also thank Marion Smedberg and Sabine Ogier-Collin for their support with the \verb|GENE-X| code. The simulations presented herein were carried out in part on the Cobra and Raven supercomputers at the Max Planck Computing and Data Facility (MPCDF) and in part on the CINECA Marconi supercomputer under the TSVVT423 project.

This work has been carried out within the framework of the EUROfusion Consortium, funded by the European Union via the Euratom Research and Training Programme (Grant Agreement No 101052200 — EUROfusion). Views and opinions expressed are however those of the author(s) only and do not necessarily reflect those of the European Union or the European Commission. Neither the European Union nor the European Commission can be held responsible for them.

%% The Appendices part is started with the command \appendix;
%% appendix sections are then done as normal sections

\appendix

\section{Normalization}
\label{appendix:normalization}

In the \verb|GENE-X| code, physical quantities are normalized to the reference density $\nref$, temperature $\Tref$, length $\Lref$, mass $\mref$, sound speed $\csref = \sqrt{\Tref / \mref}$, magnetic field $\Bref$, and elementary charge $e$. The reference sound Larmor radius is denoted by $\rhoref = \csref / \Omegaref$, and the reference plasma beta is given by $\betaref = 8 \pi \nref \Tref / \Bref^2$. This implies that the time is expressed as $t = \Lref / \csref \hat t$, the mass as $\mref \hat m_\alpha$, the temperatures as $T_\a =  \Tref \hat T_\a$ and $\tau_\a = \Tref \hat \tau_\a$, the charge as $q_\alpha = e \hat q_\alpha$, and magnetic field strength as $B = \Bref \hat B$. The spatial coordinates are normalized as $R =  \Lref \hat R$ and $Z =  \Lref \hat Z$. The generalized potential, $\psi_{1 \alpha}$, is normalized as  $\psi_{1 \alpha} = \Tref \hat \psi_{1 \alpha}  / e$, and the parallel vector potential, $A_{1 \parallel}$, is as  $A_{1 \parallel} = \rhoref \Bref \hat A_{1 \parallel}$. Finally, the spectral coefficients defined in Eq. (\ref{eq:npj}) are normalized to the reference density $\nref$ such that $\N^{pj} = \nref \nN^{pj}$. Here, quantities with a hat denote normalized values.

\section{Differential operators}
\label{appendix:differentialoperators}

The differential operators, which appear on the right-hand side of the spectral GK Vlasov equation given in Eq. (\ref{eq:spectralvlasov}), can be expressed in terms of the Poisson bracket and curvature operators, defined by

\begin{subequations} \label{eq:diffops}
  \begin{align}
    \pb{f}{g} = \bm b \times \grad f \cdot \grad g, \label{eq:pb}\\
    \curv{f} = \grad \times \bm b \cdot \grad f \label{eq:curv},
  \end{align}
\end{subequations}
\\
respectively. Here, $f$ and $g$ are two arbitrary functions. Eq. (\ref{eq:diffops}) implies that the parallel gradient operator, $\grad_{1 \parallel}$, can be written as

\begin{align} \label{eq:parallelgradient}
  \grad_{1 \parallel} f = \bm b \cdot \grad f  + \pb{f}{A_{1 \parallel}}.
\end{align}
\\
The analytical expressions of the Poisson bracket, Eq. (\ref{eq:pb}), curvature, Eq. (\ref{eq:curv}), and parallel gradient, Eq. (\ref{eq:parallelgradient}), operators written in the locally field-aligned coordinate system are given in Ref. \cite{michels2021}. With Eqs. (\ref{eq:diffops}) and (\ref{eq:parallelgradient}), the expressions of the terms involving the gradients, $\grad B$ and $\grad \psi_{ 1 \a}$, in the right-hand-side of Eq. (\ref{eq:spectralvlasov}) can be written in a form suitable to be numerically implemented using the FCI approach.

Finally, the remaining terms, associated with the divergence of the generalized fluxes in Eq. (\ref{eq:generalizedfluxes}), can be developed using the Poisson bracket and curvature operators, such that

\begin{subequations} \label{eq:divergenceidentities}
  \begin{align}
    \grad \cdot \left( \grad \times \bm b f \right)& = \curv{f} , \\
    \grad \cdot \left( \frac{\bm b \times \grad f }{B} g \right)  & =\frac{1}{B} \pb{f}{g} + \frac{g}{B} \curv{f} + \frac{g}{B^2} \pb{B}{f}.
    \end{align}
\end{subequations}
\\
With Eqs. (\ref{eq:diffops}) and (\ref{eq:divergenceidentities}), the divergence of the generalized fluxes appearing in Eq. (\ref{eq:spectralvlasov}) can be obtained in terms of the locally field-aligned coordinates.

\section{Hermite-Laguerre polynomial velocity-space basis}
\label{appendix:properties}

This appendix introduces the definitions and describe the properties of the scaled Hermite-Laguerre polynomial basis used in Eq. (\ref{eq:faexpansion}) to obtain the spectral formulation of the edge and SOL GK turbulence model presented in Section \ref{sec:spectralgenex}.

The Hermite polynomials, $H_p(x)$, of order $p$ in Eq. (\ref{eq:faexpansion}) are defined by \cite{gradshteyn2014}

\begin{align}
H_p(x) & = (-1)^p e^{x^2} \frac{d^p}{dx^p} e^{- x^2}.
\end{align}
\\
These polynomials are orthogonal over the domain $x \in \mathbb{R}$, such that 

\begin{align} \label{eq:orthohermite}
\int_{- \infty}^\infty dx H_p(x) H_{p'} e^{- x^2}& = \pi 2^p p!  \delta_p^{p’}.
\end{align}
\\
The Laguerre polynomials, $L_j(y)$, of order $j$ appearing in Eq. (\ref{eq:faexpansion}), are defined by \cite{gradshteyn2014}

\begin{align}
L_j(y) & = \frac{e^y}{j!} \frac{d^j}{dy^j} \left( e^{- y} y^j \right),
\end{align}
\\
and are orthogonal over the semi-infinite interval $x \in \mathbb{R}^+$,

\begin{align}\label{eq:ortholaguerre}
  \int_0^\infty d y L_j(y) L_{j'}(y) e^{- y} = \delta_j^{j'}.
\end{align}
\\
The Hermite-Laguerre basis form a polynomials orthogonal basis of the space of functions $f \in L^2( \mathbb{R } \times \mathbb{R}^+, e^{-x^2 - y}) $, such that

\begin{align} \label{eq:functionspace}
\int_{- \infty}^\infty dx  \int_0^\infty d y \left| f \right|^2 e^{-x^2 - y}  < + \infty,
\end{align}
\\
which is satisfied for the gyrocenter distribution function $f_\alpha$.

In order for the spectral expansion presented in Eq. (\ref{eq:faexpansion}) to provide an accurate approximation of $f_\a$, it is necessary to ensure that the amplitude of the spectral coefficients decreases sufficiently. This ensures that truncating the expansion at a sufficiently high value for both $N_{v_\parallel}$ and $N_{\mu}$ does not significantly affect the physical results. Moreover, the spectral expansion must also converge over the full range of temperatures present in the edge and SOL regions. Therefore, the width of the weight function, $F_{\mathcal{M \alpha}}$, must be appropriately defined, which is determined by the scaled temperature $\tau_\alpha$. In the present spectral approach, we consider $\tau_\a$ as a free parameter that can be chosen arbitrarily to scale the velocity-space polynomial basis to ensure an optimal convergence. One might reasonably choose to set $\tau_\alpha$ equal to the local temperature. In the this case, the spectral expansion in Eq. (\ref{eq:faexpansion}) is orthogonal to a local Maxwellian distribution function, such that the coefficients in the expansion are all zero for $p, j > 0$, if the distribution function $f_\alpha$ is given by a local Maxwellian distribution function.

While using the local temperature for $\tau_\alpha$ would enable an optimal convergence locally, it however leads to the appearance of explicit time-derivatives and gradients of the basis function in the GK Vlasov equation, given in Eq. (\ref{eq:spectralvlasov}). The presence of these terms significantly complicates the numerical implementation of the spectral formulation in \verb|GENE-X| as they introduce stiff terms that are challenging to integrate and result in asymmetric and large velocity-space spectral stencils (see Fig. \ref{fig:grid}). It should be noted that time derivatives and gradients of the fluid parallel velocity may also emerge if the velocity-space coordinates are shifted by $u_{\parallel \alpha}$ (see Eq. (\ref{eq:upara})), an approach used in Ref. \cite{frei2023b}. We remark that using the local fluid temperature is the common choice of gyromoments/gyrofluid models (see, e.g., Refs. \cite{madsen2013,frei2020}) based on Hermite-Laguerre expansion. In addition, these works consider anisotropic temperatures in the Maxwellian weight. Hence, in contrast to previous gyro-moments/fluid, we consider $\tau_\alpha$ as a constant parameter. We remark that the present discussion is restricted to the isotropic case, such that $F_{\mathcal{M} \alpha}$ is assumed to have the same width in the parallel and perpendicular direction in velocity-space, both defined by $\tau_\alpha$. However, the observation made in this section can be easily generalized to the anisotropic case. 
 
A minimal requirement for $\tau_\a$ is that it has to be chosen such that the spectral expansion of Eq. (\ref{eq:faexpansion}) can represent a Maxwellian distribution with a temperature $T_\a$, ranging from the maximal to minimal temperature in the simulations. Typically, the range of possible temperatures is determined by the Dirichlet boundary conditions imposed by the initial profiles (see Fig. \ref{fig:initialprofiles}). To derive a criterion for $\tau_\a$, we therefore assume $f_a$ is a Maxwellian distribution function with isotropic temperature $T_\a$,i.e.

\begin{align} \label{eq:famaxw}
f_a = \frac{n_\a}{\pi^{3/2} v_{T \a}^3} e^{- v_\parallel^2 / v_{T \a}^2 - \mu B / T_\a},
\end{align}
\\
with $v_{T \a}^2 = 2 T_\a / m_\a$. Using Eq. (\ref{eq:famaxw}) into Eq. (\ref{eq:npj}), we derive an explicit analytical expression for the spectral coefficients $\N^{pj}$ associated with a local Maxwellian distribution function with isotropic temperature $T_\a$, such that

\begin{align} \label{eq:npjmaxw}
\N^{pj} = n_a \left( 1 - \frac{T_\a}{\tau_\a}\right)^{\lfloor p/2 \rfloor + j} \frac{H_p(0)}{\sqrt{2^p p!}},
\end{align}
\\
where $H_p(0)$ is the Hermite number defined by \cite{gradshteyn2014}

\begin{align}
H_p(0) =  \frac{p! (-2)^{\frac{p}{2}}}{2^{p/ 2}  (p/2)! } ,
\end{align}
\\
if $p$ is even and $0$ otherwise. We remark that only the state variables contribute to Eq. (\ref{eq:npjmaxw}) since all the flux variables ($p$ odd) are identically $0$. Eq. (\ref{eq:npjmaxw}) is used as initial and Dirichlet boundary conditions of the spectral coefficients, as discussed in Section \ref{sec:numericalimplementation}. 

For the amplitude of the spectral coefficient given in Eq. (\ref{eq:npjmaxw}) to decrease with $p$ and $j$ and, thus, the spectral expansion to converge, Eq. (\ref{eq:npjmaxw}) imposes that

\begin{align} \label{eq:crittaua}
\tau_{\a c} = \frac{1}{2}\max \left(T_\a\right) <  \tau_\a.
\end{align}
\\
Eq. (\ref{eq:crittaua}) determines the maximal width of a Maxwellian distribution function that can be represented by the spectral expansion in Eq. (\ref{eq:faexpansion}) and sets, thus, a lower bound of the value of $\tau_\alpha$. As a consequence, a Maxwellian distribution function with a temperature $T_\alpha > 2 \tau_\alpha$ cannot be represented by the spectral expansion in Eq. (\ref{eq:faexpansion}) as Eq. (\ref{eq:npjmaxw}) would otherwise divergence. Therefore, to ensure that Eq. (\ref{eq:crittaua}) is satisfied initially in the simulations, we estimate $\tau_{\a c}$ with the value of the initial temperature profile (see Fig. \ref{fig:initialprofiles}) at the innermost closed flux surface. Therefore, given the initial temperature profiles, we choose $\tau_\a$ for each species to be of the order (but larger) than $\tau_{\a c}$. We remark that values of $\tau_\a$ close to (but larger than) $\tau_{\a c}$ ensure a faster convergence in the SOL region but a slower convergence in the edge region. Values of $\tau_\a$ much larger than $\tau_{\a c}$ can efficiently describe the larger temperatures in the edge, but deteriorates the spectral convergence in the SOL. 

In practice, we determine the most appropriate choice of $\tau_\a$ by trial and error given in the initial temperature profile to ensure stable numerical simulations. We find that, in general, choosing $\tau_\a \gtrsim \tau_{\a c}$ is a suitable choice in the case of the TCV-X21 spectral simulations (see Fig. \ref{fig:initialprofiles}).

%% If you have bib database file and want bibtex to generate the
%% bibitems, please use
%%
\bibliographystyle{elsarticle-num} 
\bibliography{library}

\end{document}